\documentclass[]{interact}

\usepackage{epstopdf}
\usepackage[caption=false]{subfig}
\usepackage[authoryear]{natbib}
\renewcommand\bibfont{\fontsize{10}{12}\selectfont}
\makeatletter
\def\NAT@def@citea{\def\@citea{\NAT@separator}}
\makeatother

\theoremstyle{plain}

\theoremstyle{definition}

\theoremstyle{remark}

\usepackage{adjustbox} 

\usepackage{bm}
\usepackage{booktabs}
\usepackage{amsmath,amsfonts,amssymb,amsthm}
\usepackage{graphicx}
\usepackage{enumerate}

\begin{document}


\title{On Extreme Value Index Estimation under Random Censoring}

\author{\name{Richard Minkah\textsuperscript{a}\thanks{CONTACT R. Minkah. Email: rminkah@ug.edu.gh} and\\ Tertius de Wet\textsuperscript{b} and\\ Kwabena Doku-Amponsah\textsuperscript{a}}\affil{\textsuperscript{a}Department of Statistics and Actuarial Science,\\ University of Ghana, Ghana; \textsuperscript{b}Department of Statistics and Actuarial Science,\\ Stellenbosch University,\\ South Africa}}

\maketitle

\begin{abstract}
	Extreme value analysis in the presence of censoring is receiving much attention as it has applications in many disciplines, including survival and reliability studies. Estimation of extreme value index (EVI) is of primary importance as it is a critical parameter needed in estimating extreme events such as quantiles and exceedance probabilities. In this paper, we review several estimators of the extreme value index when data is subject to random censoring. In addition, four  estimators are proposed, one based on the exponential regression approximation of log spacings, one based on a Zipf estimator and two based on variants of the moment estimator. The proposed estimators and the existing ones are compared under the same simulation conditions. The performance measures for the estimators include confidence interval length and coverage probability. The simulation results show that no estimator is universally the best as the estimators depend on the size of the EVI parameter, percentage of censoring in the right tail and the underlying distribution. However, certain estimators such as the proposed reduced-bias estimator and the adapted moment estimator are found to perform well across most scenarios. Moreover, we present a bootstrap algorithm for obtaining samples for extreme value analysis in the context of censoring. Some of the estimators that performed well in the simulation study are illustrated using a practical dataset from medical research

\end{abstract}


\section{Introduction}	

Statistics of extremes under random censoring is a relatively new area in extreme value analysis that has received considerable attention in the literature during the last few years. Examples of applications include estimating survival time \citep{Einmahl2008,Ndao2014} and large insurance claims \citep{Beirlant2017}, among others. In order to obtain estimates of parameters of extreme events, the extreme value index (EVI) is the primary parameter needed. Although the EVI estimation in the case of complete samples has been studied extensively, the same cannot be said of the censored case. In this paper, we review existing estimators and propose two estimators that are aimed at reducing the bias and variance. In addition, we provide a simulation comparison of the various estimators of the Extreme Value Index (EVI).    

The first work on the subject can be attributed to \citet{Beirlant2001}. The authors proposed an adaptation of the Hill estimator under  random right censoring. The motivation for this adapted Hill estimator to censoring was the same as that of the Hill estimator obtained from the slope of the Pareto quantile plot. However, since the censored observations have the same values (i.e the maximum), the Pareto quantile plot will be horizontal in those observations. As a result, the adaptation of the Hill estimator to censoring was based on the slope of the Pareto quantile plot for the noncensored observations only. In addition, by using the second order properties of the representation of log-spacings in the exponential regression model, a bias-corrected version of the adapted Hill estimator was obtained. The finite sample properties of the estimator were studied through a simulation study and the estimator was found to give credible estimates for a percentage of censoring of 5\% at most. Consistency and asymptotic normality of the estimator were obtained under some restrictive conditions on the number of noncensored observations and the sample tail fraction. \citet{Delafosse2002} proved the almost sure convergence of the adapted Hill estimator in \cite{Beirlant2001} under very general conditions on the number of noncensored observations.

Also, in \citet[Section 6.1]{Reiss2007}, 
the authors introduced an estimator of the EVI when data is  randomly-or fixed censored. In the case of random right censoring, the Pareto or generalised Pareto distribution was fitted to the excesses over a given threshold. The likelihood function of the chosen distribution was adapted to censoring and maximised to obtain an estimator of the EVI. However, the authors made no attempt  to study the asymptotic properties of the their proposed estimators of the EVI. 

In addition, \citet{Beirlant2007} proposed an entirely different approach by adapting the estimator of the EVI from the Peaks-Over Threshold (POT) method \citep{Smith1987} and the moment estimator \citep{Dekkers1989} to random right censoring. The former estimator involved adapting the likelihood function to the context of censoring whereas the latter estimator was obtained by dividing the classical EVI estimator by the proportion of noncensored observations in the top order statistics selected from the sample. Due to the difficulties in establishing the asymptotic properties of the maximum likelihood estimator of the POT method, \citet{Beirlant2010} proposed a one-step approximation based on the Newton-Raphson algorithm. The reported simulation study showed the closeness of the approximation of the one-step estimators to the maximum likelihood estimators. The added advantage was that the asymptotic normality of the one-step estimators has been established, unlike that for the maximum likelihood estimators.

Based on the ideas of \citet{Beirlant2007}, \citet{Einmahl2008} provided a second methodological paper which considered estimators based on the top order statistics. In addition, the authors proposed a unified method to prove the asymptotic normality of the EVI estimators. A small scale simulation showed the superiority of the adapted Hill estimator for the Pareto domain of attraction and a slight advantage of the adapted generalised Hill for the Weibull and the Gumbel domains of attraction. \citet{Einmahl2008} used restrictive conditions to prove the asymptotic normality of the EVI estimators. However, these conditions were relaxed by \citet{Brahimi2013} to prove the asymptotic normality of the adapted Hill estimator of the EVI under  random right censoring.   

The estimation of the EVI has also received attention from \citet{Gomes2010} and \citet{Gomes2011}. These papers form an overview of the EVI estimators in the context of random censoring. To the best of our knowledge, \citet{Gomes2011} made the first attempt at introducing a reduced-bias estimator of the EVI, in the form of the minimum-variance reduced-bias (MVRB) estimator \citep{Caeiro2005}. The reported simulation study showed an overall best performance for the adapted MVRB estimator for samples generated from distributions from the Pareto domain of attraction. As in \citet{Einmahl2008}, the generalised Hill performed better than the other adapted EVI estimators for samples whose underlying distribution functions are in the Weibull and Gumbel domains of attraction.

The Hill estimator for estimating the EVI under random censoring performs well, although in the classical case it is known to be biased, not location invariant and unstable. Efforts have been made to provide reduced-bias and minimum variance Hill-type estimators to improve on the Hill estimator for the heavy-tailed distributions (i.e. distributions in the Pareto domain of attraction). In this regard, \citet{Worms2014} provided another methodological paper for the estimation of the EVI in the case of censoring. They provided two sets of Hill-type estimators based on the Kaplan-Meier estimation of the survival function \citep[see][]{Kaplan1958} and the synthetic data approach of \citet{Leurgans1987}. In addition, the authors presented a small scale simulation that compared the performance of the two proposed estimators to the adapted Hill and MVRB estimators. The results showed that the two proposed estimators are superior to the Hill estimator, in particular, the estimator based on the ideas of \citet{Leurgans1987}. On the other hand, MVRB performed better than the authors' proposed estimators. However, the EVI estimator based on the synthetic data approach of \citet{Leurgans1987} compared favourably in the strong censoring framework with the MVRB estimator. The consistency of these estimators was proved under mild censoring. However, the asymptotic normality of these two estimators remains an open problem.

Furthermore, the estimation of the EVI for the Pareto domain of attraction has also been obtained from the Bayesian perspective by \citet{Ameraoui2016}. They constructed a maximum aposteriori and mean posterior estimators for various prior distributions of the EVI, namely Jeffrey’s, Maximal Data Information (MDI) and a conjugate Gamma. The asymptotic properties, namely consistency and normality of the estimators,  were established. A small simulation study was used to examine the finite sample properties and the performance of the estimators. The reported simulation result showed the superiority of the maximum aposteriori estimator under maximal data information prior. 

We aim to achieve two objectives in this paper. Firstly, we propose some estimators of the EVI including a reduced-bias estimator based on the exponential regression model of \citet{Beirlant1999a}. Secondly, the above researchers compared their proposed estimators under different simulation conditions. In addition, some of the estimators' asymptotic distributions remain an open problem, and hence, theoretical comparison is not possible. Therefore, the second objective of this paper is to compare several of the existing estimators with the proposed ones in a simulation study under identical conditions. 

The rest of the paper is organised as follows. In Section \ref{sec2}, we present the framework of extreme value analysis when data is censored. In Section \ref{sec_sim}, a simulation comparison of the various estimators is presented. In Section \ref{sec_prac}, we present a practical application of the estimators to estimate the extreme value index for a medical data set on the survival of AIDS patients. Lastly, concluding remarks are presented in Section \ref{sec_conc}.

\section{Framework}	\label{sec2}

Let $X_1,X_2,..., X_n$ be a sequence of independent and identically distributed (\textit{i.i.d}) random variables with distribution function $F,$ and $X_{1,n} \leq X_{2,n}\leq ... \leq X_{n,n}$ the associated order statistics. Therefore, the sample maximum is denoted by $X_{n,n}$. Extreme value theory attempts to solve the problem of the possible limit distributions of $X_{n,n}$. It is well-known that the distribution of the sample maximum can be obtained from the underlying distribution of $X$ as 

\begin{equation}\label{max}
F_{X_{n,n}}(x)=F^n(x).
\end{equation}
However, $F$ is usually unknown and, hence, EVT focuses on the search for an approximate family of models for $F^n$ as $n\to\infty.$  

Limiting results for  $F^n$ in EVT have been addressed in the papers by 
\citet{Fisher1928} and \citet{Gnedenko1943}. Specifically, the results can be stated as follows: if there exist sequences of constants  $b_n$ and $a_n>0$ $(n=1, 2,...)$, such that \begin{equation} \label{LimDist}
\lim\limits_{x \to \infty}P\left(\frac{X_{n,n}-b_n}{a_n}\leq x\right)\rightarrow \Psi(x), 
\end{equation}
where $\Psi$ is a nondegenerate distribution function, then $\Psi$ belongs to the  family of distributions,

\begin{equation} \label{GEV}
\Psi_\gamma(x)=\left\{\begin{array}{ll}
\exp\left(-\left(1+\gamma\frac{x-\mu}{\sigma}\right)^{-1/\gamma}\right), & 1+\gamma\frac{x-\mu}{\sigma}>0,~ \gamma\ne 0,\\
\exp\left(-\exp\left(\frac{x-\mu}{\sigma}\right)\right), & x\in\mathbb{R}, ~\gamma=0,\end{array} \right.
\end{equation}
where $\mu \in\mathbb{R} $ and $\sigma>0.$ The quantity $\gamma\in\mathbb{R}$, is the \emph{Extreme Value Index} (EVI) or the \emph{tail index}: it determines the tail heaviness of the extreme value distributions. The EVI is classified into three groups, each representing one of the three families of distributions, Gumbel (exponential tails), Pareto (heavy-tailed) and Weibull (light-tailed). The group of families have $\gamma=0,$ $\gamma>0$ and $\gamma<0$ corresponding to the Gumbel, Pareto and Weibull families respectively.  A distribution function $F$ satisfying (\ref{GEV}) is said to be in the maximum domain of attraction of $\Psi_\gamma$ written as $F\in D\left(\Psi_\gamma\right).$

In addition to (\ref{GEV}), \citet{Balkema1974} and \citet{Pickands1975} showed the generalised Pareto distribution (GPD) as the limit distribution of scaled excesses over a sufficiently large threshold. The GPD can be written as

\begin{equation} \label{GPD}
\Lambda_{\gamma}(x)= 1+\ln\Psi_\gamma(x)=\left\{\begin{array}{ll}
1-\left(1+\gamma\frac{x-\mu}{\sigma}\right)^{-1/\gamma}, & 1+\gamma\frac{x-\mu}{\sigma}>0,~ \gamma\ne 0,\\
1-\exp\left(\frac{x-\mu}{\sigma}\right), & x\in\mathbb{R}, ~\gamma=0,\end{array} \right.
\end{equation}
where $\Psi_\gamma$ is given in (\ref{GEV}).

In this paper, our interest is in the Pareto domain of attraction i.e. the case $\gamma>0.$ This family consists of distribution functions $F$ whose tails are regularly varying with a negative index of variation. That is

\begin{equation}\label{FtDomain 1-F}
1-F(x)=x^{-1/\gamma}\ell_F(x), ~~x\to \infty,
\end{equation}
where $\ell_F$ is the slowly varying function associated with $F.$ A slowly varying function, $\ell,$  is of the form $\ell(xt)/\ell(x)\to 1$ for $x\to\infty.$ Relation (\ref{FtDomain 1-F}) can be stated equivalently in terms of the associated upper tail quantile function $U$ as
\begin{equation}\label{FtDomain U}
U(x)=F^{-1}(1-\frac{1}{x})=x^{\gamma}\ell_U(x),~~ x\to \infty,
\end{equation}
where  $\ell_U$ is the slowly varying function associated with $U$.

\subsection{EVT Conditions}	
The conditions underlying domain of attraction  are presented in this section. These conditions are needed in defining estimators of tail parameters and to study their asymptotic properties. 

\citet{deHaan1984} gave the following well-known necessary and sufficient condition for $F\in D(\Psi_\gamma),$ known as the first-order condition or extended regular variation: 

\begin{equation}\label{cond1}
\lim\limits_{u \to \infty}\frac{U(ux)-U(u)}{a(u)}=h_\gamma(x):=\left\{ \begin{array}{ll}
\frac{x^\gamma-1}{\gamma} & \mbox{if $\gamma\ne 0$}\\
\log{x} & \mbox{if $\gamma= 0$},\end{array} \right.
\end{equation}
where $a$ is a positive measurable function, $x>0.$

In addition, to study the asymptotic properties of the estimators of tail parameters, the first-order condition is generally not sufficient; a second-order condition specifying the rate of convergence of (\ref{cond1}) is also required. 


In the literature, the second-order condition can be stated in terms of $U$ \citep[see e.g.][]{deHaan2006,Gomes2008}, or, equivalently, also in terms of the rate of convergence of the slowly varying function, $\ell,$ in (\ref{FtDomain U}). \citet[page  602]{Beirlant1999a} state it as follows:
there exists a real constant $\rho<0$ and a rate function $b$ satisfying $b(x)\to 0$ as $x\to \infty,$ such that for all $\lambda\geq 1$, 

\begin{equation}\label{2nd s.v}
\lim\limits_{x\to \infty}\frac{\log\ell(\lambda x)-\log\ell(x)}{b(x)}=\kappa_\rho(\lambda),
\end{equation}
where $\kappa_\rho(\lambda)=\int_{1}^{\lambda}u^{\rho-1}du.$

\subsection{General Estimation under Censored Data}\label{EVTCEN_genEst}

Let the random variable of interest be $X$ with distribution function, $F.$ Since samples on $X$ may not be fully observed, we introduce another positive random variable $C,$ which is independent of $X,$ with distribution function $G.$ In this setting, we then observe $\left(Z_i,\delta_i\right), i=1,\ldots,n$ with

\begin{equation} \label{Zi}
Z_i=\mbox{min}\left(X_i, C_i\right)
\end{equation}

and 

\begin{equation}
\delta_i=\left\{ \begin{array}{ll}
1 & \mbox{if $X_i \leq C_i$};\\
0 & \mbox{if $X_i > C_i$}.\end{array} \right.
\end{equation}
Here, $\delta_i$ is a variable indicating whether $Z_i$ is censored or not. Let $H$ be the distribution function of $Z$ defined in (\ref{Zi}). Thus, by the independent assumption of the random variables $Y$ and $C,$ we have $1-H=(1-F)(1-G).$

In addition, let $\vartheta_F=\sup{\{F(x)<1\}}$ be the corresponding right endpoint of the underlying distribution function, $F.$ Similarly, let $\vartheta_G~ \mbox{and}~ \vartheta_H$ be the right endpoints of the underlying distribution functions of $C$ and $Z$ respectively. If we assume $F\in D(\Psi_{\gamma_1})$ and $G\in D(\Psi_{\gamma_2})$ for some real numbers, $\gamma_1~\mbox{and}~\gamma_2,$ then $H\in D(\Psi_{\gamma})$ where $\gamma \in \mathbb{R}.$ \citet{Einmahl2008} considered these three  combinations of $\gamma_1$ and $\gamma_2:$

\begin{enumerate}[\hspace{0.5cm}\textbf{Case} \bfseries 1.]
	\item  $ F~ \mbox{and}~ G~ \mbox{are Pareto types:~~}~\gamma_1>0, ~\gamma_2>0 ~~~~~~~~~~~~~~~~~~~~~~~~~\to~ \gamma=\frac{\gamma_1\gamma_2}{\gamma_1+\gamma_2}$
	\item $ F~ \mbox{and}~ G~ \mbox{are Gumbel types:}~\gamma_1=0, ~\gamma_2=0, ~\vartheta_F=\vartheta_G ~~~~~~~\to~~\gamma=0$
	\item $ F~ \mbox{and}~ G~ \mbox{are Weibull types:} ~\gamma_1<0, ~\gamma_2<0,~ \vartheta_F=\vartheta_G =\infty~ \to ~\gamma=\frac{\gamma_1\gamma_2}{\gamma_1+\gamma_2}.$
\end{enumerate}
The other two possibilities, $\{\gamma_1>0, ~\gamma_2<0\}$  and $\{\gamma_1<0, ~\gamma_2>0\},$ correspond closely to the completely noncensored case which has been studied widely whereas the latter corresponds closely to the completely censored case where estimation is impossible. 

\subsection{Extreme Value Index Estimation Methods }\label{EVI_CEN}

The estimation of the extreme value index (EVI) when observations are censored needs some modification from that of the complete sample. This is because the observed sample is $(Z_i,\delta_i),~i=1,\ldots, n,$ and hence, the application of the classical EVI estimation methods will yield estimators that converge to $\gamma,$ the EVI of the underlying distribution of the random variable $Z.$ However, our interest is in $\gamma_1,$ the EVI of the underlying distribution of the random variable $X.$ Therefore, some modification is needed to adapt the estimation of $\gamma$ from the $Z$ sample to estimate $\gamma_1.$   

The existing methodologies for estimating the EVI under right censoring can be grouped into four categories: 

\begin{enumerate}[\hspace{0.5cm}\bfseries 1.]
	\item	adapting a classical EVI by dividing it by the proportion of noncensored observations \citep{Beirlant2007,Einmahl2008,Gomes2011}; 
	\item	 adapting the likelihood function of an extreme value distribution \citep{Beirlant2010};
	\item Censored regression \citep{Worms2014}.
	\item Bayesian estimation \citep{Ameraoui2016,Beirlant2017}
\end{enumerate}
In this paper, we consider the frequentist methods only i.e. the first three cases. The methods are grouped into three categories and presented together with the proposed estimators in the three sub-sections that follow. Following that, we propose four further estimators that are adapted to the censored case. 

\subsubsection{First Method}
The first method was introduced in \citet{Beirlant2007} and further developed by \citet{Einmahl2008}. In this method a classical estimator of the EVI is obtained from the $Z$ sample and then adapted to censoring.  Among these estimators are: the maximum likelihood estimator from the Peaks-Over Threshold (POT) method and the moment estimator \citep{Beirlant2007}; Hill, Moment, Generalised Hill and the maximum likelihood estimator from the POT method \citep{Einmahl2008}; and Hill, moment, mixed moment and generalised Hill \citep{Gomes2011}. In addition, \citet{Einmahl2008} provides a uniform way to establish the asymptotic normality of the proposed estimators of the EVI (i.e. Hill, Moment, Generalised Hill and the maximum likelihood estimators). These estimators are reviewed below in terms of the random variable $Z,$ and thus estimates $\gamma$ the EVI  $Z.$

\textbf{The Hill Estimator}: The Hill estimator \citep{Hill1975} is arguably the most common estimator of $\gamma$ in the Pareto case i.e. $\gamma>0.$ The Hill estimator is defined for the $(k+1)$-largest order statistics as

\begin{equation}\label{Hill}
\hat{\gamma}^{(Hill)}_{Z,k,n}=\frac{1}{k}\sum_{j=1}^{k}\log{Z_{n-j+1,n}}-\log{Z_{n-k,n}}.  
\end{equation}

The properties of the Hill estimator have been studied widely and its attractive properties include consistency \citep{Mason1982} and asymptotic normality \citep{Hall1982,deHaan1998}. 

\textbf{The Generalised Hill Estimator}: \citet{Beirlant1996} proposed the generalised Hill (UH) estimator in a bid to extend the Hill estimator to the case where $\gamma \in \mathbb{R}.$ The UH estimator is obtained as the slope of the ultimately linear part of the generalised Pareto quantile plot, 
\begin{equation}\label{GPQ}
\left(-\log{\left(\frac{j+1}{n+1}\right)},~ \log{(Z_{n-j,n}H_{Z,j,n})}\right), j=1,2, ..., n-1.
\end{equation}
It is  given by
\begin{equation}\label{UH}
\hat{\gamma}^{(UH)}_{Z,k,n}= \frac{1}{k}\sum_{j=1}^{k}\log{ UH}_{Z,j,n}-\log UH_{Z,k+1,n}, 
\end{equation}
where $UH_{Z,j,n}=Z_{n-j,n}\left( \frac{1}{j}\sum_{i=1}^{j}\log{Z_{n-i+1,n}}-\log{Z_{n-j,n}}\right).$

\textbf{The Minimum-Variance Reduced Bias  Estimator}: \citet{Caeiro2005}  proposed the Minimum-Variance Reduced Bias (MVRB) estimator for heavy-tailed distributions belonging to the Hall class \citep{Hall1982} of models. The estimator is a direct modification of the Hill estimator using the second order parameters to reduce bias. It has the added advantage of having the same asymptotic variance as the Hill estimator. The MVRB estimator is obtained by using the  second-order condition (\ref{2nd s.v}) with $b(u)=\gamma\beta u^\rho.$ It is given by

\begin{equation}\label{MVRB}
\hat{\gamma}^{(MVRB)}_{Z,k,n}= \hat{\gamma}^{(Hill)}_{Z,k,n} \left(1-\frac{\hat{\beta}}{1-\hat{\rho}}
\left(\frac{k}{n}\right)^{-\hat{\rho}}\right),
\end{equation}

where, $\hat{\gamma}^{(Hill)}_{Z,k,n}$ is the Hill estimator in (\ref{Hill}) and the pair  $(\hat{\beta}, \hat{\rho})$ is the estimator for the pair of parameters  $(\beta, \rho)$ of the second-order auxiliary function $b.$  

\textbf{The Moment Estimator}: \citet{Dekkers1989} introduced another estimator known as the moment estimator as an adaptation of the Hill estimator valid for all domains of attraction. The moment estimator is defined for $k\in \{2, ..., n-1\}$ and it is given by

\begin{equation}\label{MOM}
\hat{\gamma}^{(MOM)}_{Z,k,n}= M_{Z,k,n}^{(1)}+1-\frac{1}{2}\left(1-\frac{(M_{Z,k,n}^{(1)})^2}{M_{Z,k,n}^{(2)}}\right)^{-1},
\end{equation}

where
\begin{equation}\label{Mj}
M_{Z,k,n}^{(j)}=\frac{1}{k}\sum_{i=1}^{k}(\log{Z_{n-i+1,n}}-\log{Z_{n-k,n}})^j, ~ j=1,2.
\end{equation}


\subsubsection*{Adapting EVI Estimators} 
\cite{Beirlant2007} and \citet{Einmahl2008} proposed that the EVIs for the complete sample, $\gamma^{(.)}_{Z,k,n},$ (i.e. (\ref{Hill}) - (\ref{PMoM})) can be adapted to censoring by dividing each estimator by the proportion of noncensored observations, $\hat{\wp},$ in the $k$ largest $Z$ observations. Thus, the estimator of $\gamma_1$ is given by

\begin{equation}\label{adapt_Z}
\hat{\gamma}_1=\hat{\gamma}_{Z,k,n}^{(c, .)} =\frac{\hat{\gamma}_{Z,k,n}}{\hat{\wp}}.
\end{equation}
Here, $\hat{\wp}$ is given by  
\begin{equation} 
\hat{\wp}=\frac{1}{k}\sum_{i=1}^{k}\delta_{n-i+1, n},
\end{equation}
where $\delta_{i,n},~ i=1, ..., n$ are the $\delta$-values corresponding to $Z_{i,n},~ i=1, ..., n$ respectively. In the literature, (\ref{adapt_Z}) has primarily been used to adapt the EVI estimators to censoring.

\subsubsection{Second Method}

The second method introduced by \citet{Beirlant2010} involves using the POT method and adapting the log-likelihood function for censoring. We know from (\ref{GPD}) that given a high threshold, $u,$ the limit distribution of excesses  $V_j=Z_i-u, j=1, \ldots, k$  given $Z_i>u, i=1, \ldots, n$  can be approximated by the generalised Pareto (GP) distribution. In \citet{Beirlant2007} and \citet{Einmahl2008}, the maximum likelihood estimator, $\hat{\gamma}^{(c,POT)_{Z,k,n}},$ is obtained from the GP approximation of the distribution of the $V_j$'s  and is adapted to censoring using (\ref{adapt_Z}). 

An alternative approach in \citet{Beirlant2010} involves adapting the likelihood function of the random variable $V_j, j=1, \ldots,k$,  

\begin{equation}\label{c.potlik}
L(\gamma_1, \sigma_{1,k})=\Pi_{j=1}^{k}[\lambda(V_j)]^{\delta_j}[1-\Lambda(V_j)]^{1-\delta_j}
\end{equation}
where $\Lambda$ is the GP distribution and $\lambda$ the corresponding density function of the GP distribution. However, there are difficulties with obtaining explicit expressions for the maximum likelihood estimators of $\gamma_1$ and $\sigma_{1,k}.$ In addition, their asymptotic properties remain an open problem. As a result, \citet{Beirlant2010} proposed solving the maximum likelihood equations using one-step approximations based on the Newton-Raphson algorithm. The resulting estimator of the parameters is given by 

\begin{equation}\label{Newton}
\left( \begin{array}{c}
\hat{\gamma}_{Z,k,n}^{(c,POT.L)}\\\\
\frac{\hat{\sigma}_{Z,k}^{(c,POT.L)}}{\sigma_{1,k}}
\end{array}\right)  =\left( \begin{array}{c}
\hat{\gamma}_{Z,k,n}^{(c, I)}\\\\
\frac{\hat{\sigma}_{Z,k}^{(c,I)}}{\sigma_{1,k}}
\end{array}\right)-
\left(
\begin{array}{cc}
L_{11}^{''} & \sigma_{1,k}L_{12}^{''}\\\\
\sigma_{1,k}L_{12}^{''} & \sigma_{1,k}^{2}L_{22}^{'}
\end{array}
\right) \left( 
\begin{array}{c}
L_{1}^{'} \\\\
\sigma_{1,k}L_{2}^{'}
\end{array}
\right)
\end{equation} 
where $L_{i}^{'}$ and $L_{ij}^{''},~ i=1,2,j=1,2$ are the first and second derivatives of $\log{L(\gamma_1, \sigma_{1,k})},$ evaluated at $\left(\hat{\gamma}_{Z,k,n}^{(c, I)},~ \hat{\sigma}_{Z,k}^{(c,I)}\right).~$ The estimators, $\hat{\gamma}_{Z,k,n}^{(c, I)}~ \mbox{and}~\hat{\sigma}_{Z,k}^{(c,I)},$ are the initial estimators and must be asymptotically normal. The authors state that the moment estimator provides a good example of the initial estimators. The performance of the estimators, $\hat{\gamma}_{Z,k,n}^{(c,POT.L)}$ and $\hat{\sigma}_{Z,k}^{(c,POT.L)},$ were found to be close to the maximum likelihood estimators obtained from (\ref{c.potlik}). In addition, the asymptotic normality of the one-step Newton-Raphson estimators obtained in (\ref{Newton}) has been established in that paper.

\subsubsection{Third Method} 
The third method introduced by \citet{Worms2014} is based on censored regression method of \citet{Koul1981}. The estimators are valid for estimating the EVI for distributions in the Pareto domain of attraction. From the well known result of deriving the Hill estimator from the mean excess function, they define an adaptation of the classical Hill estimator valid for case 1 as,
\begin{equation}\label{WW.KL}
\hat{\gamma}_{Z,k,n}^{(c, WW.KM)} := \frac{1}{n(1-\hat{F}(Z_{n-k,n}))}\sum_{j=1}^{k}\frac{\delta_{n-j+1,n}}{1-\hat{G}(Z_{n-j+1,n}^-)}\log{\left(\frac{Z_{n-j+1,n}}{Z_{n-k,n}}\right)},
\end{equation}
where $\hat{F}$ and $\hat{G}$ are the  Kaplan-Meier estimators for $F$ and $G$ respectively. Here, the Kaplan-Meier estimators of the survival functions are defined for $b<Z_{n,n}$ as

\begin{equation}
1-\hat{F}(b)= \Pi_{Z_{j,n}\leq b} \left( \frac{n-j}{n-j+1}\right)^{\delta_{j,n}}
\end{equation}
and
\begin{equation}\label{1-G(Z^-)}
1-\hat{G}(b)= \Pi_{Z_{j,n}\leq b} \left( \frac{n-j}{n-j+1}\right)^{1-\delta_{j,n}}.
\end{equation}
In practice, the estimator ${1-\hat{G}(Z_{n-j+1,n}^-)}$ can be equal to zero, making (\ref{WW.KL}) undefined. Therefore, \citet{Worms2014} defined $\hat{G}(Z_{n-j+1,n}^-)$ as a function of the form  $g(z^-)=\lim\limits_{\nu\to z}g(\nu).$ 

As an alternative to the Kaplan-Meier estimators of $F$ and $G,$ \citet{Worms2014} provides a variant of (\ref{WW.KL}) based on the ideas of ``synthetic data" introduced by \citet{Leurgans1987}. The estimator turns out to be a weighted version of the Hill estimator, (\ref{WW.KL}), and is given by

\begin{equation}\label{WW.L}
\hat{\gamma}_{Z,k,n}^{(c, WW.L)} := \frac{1}{n(1-\hat{F}(Z_{n-k,n}))}\sum_{j=1}^{k}\frac{\delta_{n-j+1,n}}{1-\hat{G}(Z_{n-j+1,n}^-)}j\log{\left(\frac{Z_{n-j+1,n}}{Z_{n-k,n}}\right)}.
\end{equation}
The consistency of the estimators (\ref{WW.KL}) and (\ref{WW.L}) were proven under some restrictive conditions. However, the asymptotic normality of the estimators (\ref{WW.KL}) and (\ref{WW.L}) remains an open-problem.

\subsubsection{The Proposed Estimators}

We propose adapting the exponential regression method of \citet{Beirlant1999a} to censoring. This method yields a  maximum likelihood (ML) estimator for $\gamma>0,$ and hence, for $\gamma_1>0.$ 

\citet{Beirlant1999a}  provide an approximate representation for the log-spacings of successive order statistics:  

\begin{equation}\label{L-spacings}
R_j=j(\log{Z_{n-j+1,n}}-\log{Z_{n-j,n}})\sim \left( \gamma+b_{n,k}\left(\frac{j}{k+1}\right)^{-\rho} \right)E_j,  ~j=1, \ldots, k,
\end{equation}
where $E_j, ~j=1, ..., k$ are standard exponential random variables, $b_{n,k}=b\left((n+1)/(k+1)\right)\in \mathbb{R}~ (\mbox{also}~ b_{n,k}\to 0,~\mbox{as}~k,n\to \infty)$ and $\rho$ are second-order parameters from (\ref{2nd s.v}). From the approximate distribution of log-spacings (\ref{L-spacings}), a likelihood function can be formed. Maximisation of the likelihood function leads to the maximum likelihood estimators $\hat{\gamma}_{Z,k,n}^{(ERM)}, \hat{b}_{n,k}~ \mbox{and}~ \hat{\rho}$ of $\gamma, ~b_{n,k}~\mbox{and}~ \rho$ respectively. We note that (\ref{L-spacings}) simplifies to $R_j \sim \gamma E_j, ~ j=1, ..., k$ if $ b_{n,k}=0.$ In addition, the resulting maximum likelihood estimator is the usual Hill estimator. 

The maximum likelihood estimator, $\hat{\gamma}_{Z,k,n}^{(ERM)},$ of $\gamma$ is adapted to censoring to obtain an estimator of $\gamma_1$ using (\ref{adapt_Z}). Moreover, the estimation of $\gamma$ leads to concurrent estimates of the second order parameters, $\hat{b}_{n,k}~ \mbox{and}~ \hat{\rho}.$ These estimators can be adapted to censoring and used to obtain reduced-bias estimators for quantiles and exceedance probabilities.

In addition, we propose adapting the Zipf estimator of
\citet{Kratz1996}. This estimator is a smoother version of the Hill estimator and is is valid for $\gamma>0.$ The estimator is obtained through a minimisation of the unconstrained least squares function involving the $k$ largest observations on the generalised Pareto quantile (\ref{GPQ}),
\[
L(\gamma, \eta)=\sum_{j=1}^{k}\left(\log Z_{n-j,n}H_{Z,j,n}-\eta+\gamma\log{\frac{j+1}{n+1}}\right)^2,
\]
with respect to $\eta$ and $\gamma.$  This results in the Zipf estimator given by

\begin{equation}\label{Zipf}
\hat{\gamma}^{(Zipf)}_{Z,k,n}= \frac{\frac{1}{k}\left(\sum_{j=1}^{k}\log{\frac{k+1}{j+1}}-\frac{1}{k}\sum_{i=1}^{k}\log{\frac{k+1}{i+1}}\right)\log {Z_{n-j+1,n}}}{\frac{1}{k}\sum_{j=1}^{k}\log^2{\frac{k+1}{j+1}-\left(\frac{1}{k}\sum_{i=1}^{k}\log{\frac{k+1}{j+1}}\right)^2}}. 
\end{equation}
The estimator, $\hat{\gamma}^{(Zipf)}_{Z,k,n},$ in (\ref{Zipf}) is adapted to censoring, $\hat{\gamma}^{(c,Zipf)}_{Z,k,n},$ using (\ref{adapt_Z}).

Furthermore, the popularity of the moment estimator, (\ref{MOM}), has led to the development of a couple of variants to deal with its shortcomings. In the case where there is no censoring, the moment ratio \citep{Danielsson1996} and the Peng's Moment \citep{Deheuvels1997} are examples of these estimators. We present these estimators and propose its adaptation to the case where observations are subject to right random censoring.	

The moment ratio estimator unlike the moment estimator,(\ref{MOM}), is valid for the Pareto domain of attraction only. It is given by

\begin{equation}\label{MoMR}
\hat{\gamma}^{(MomR)}_{Z,k,n}=\frac{1}{2}\frac{M_{Z,k,n}^{(2)}}{M_{Z,k,n}^{(1)}}.
\end{equation}
where $M_{Z,k,n}^{(j)},~j=1,2$ is defined in (\ref{Mj}). The moment ratio estimator has been shown to have a smaller asymptotic bias than the Hill estimator and a moderate mean square error at the same value of $k$ \citep{Danielsson1996}. 

The Peng's Moment Estimators is designed to reduce bias in the moment estimator and it is given by
\begin{equation}\label{PMoM}
\hat{\gamma}^{(PMom)}_{Z,k,n}= \frac{1}{2}\frac{M_{Z,k,n}^{(2)}}{M_{Z,k,n}^{(1)}}+1-\frac{1}{2}\left(1-\frac{(M_{Z,k,n}^{(1)})^2}{M_{Z,k,n}^{(2)}}\right)^{-1}.
\end{equation}
This estimator is valid for all domains of attraction and was shown to be asymptotically normal under appropriate conditions on $k.$

In the case where observations are subject to censoring, we also adapt the estimators (\ref{MoMR}) and (\ref{PMoM}) to censoring using (\ref{adapt_Z}).

\section{Simulation Study}\label{sec_sim}

To investigate and compare the performance of different EVI estimators, we shall make use of simulation. The simulation study is grouped into two categories: point and confidence interval estimation. The former involves assessing the performance of the estimators in terms of Median Absolute Deviation (MAD) and median bias. The latter case consists of diagnostic checks on 95\% confidence intervals based on the coverage probabilities and interval lengths. 

We consider the following combination of factors in the simulation: distributions, sample sizes, threshold levels, proportions of censoring. Several samples sizes, $n=500, 1000, 2000$ and 5000, and number of top order statistics, taken as 10\%, 20\% and 30\% of the sample size. However, the result did not differ so much and hence, for brevity and ease of presentation, we consider samples of size, $n=1000$ and the number of top order statistics taken as 10\% of the sample size.  

Data were generated from the three distributions presented in Table \ref{Dist}.
\begin{table}[htp!]
	\centering
	\caption{Distributions}
	\begin{tabular}{ccc}
		\toprule
		Distribution         &                               $1-F(z)$                                &        $\gamma$         \\ \hline
		&                                                                       &                         \\
		Burr ($\eta, \tau, \lambda$) & $\left(\eta/(\eta+z^\tau)\right)^\lambda,~~~z>0;\eta,\lambda,\tau>0 $ & $\frac{1}{\tau\lambda}$ \\
		&                                                                       &                         \\
		Pareto ($\alpha$)       &                     $z^{-\alpha},~~~z>1;\alpha>0$                     &   $\frac{1}{\alpha}$    \\
		&                                                                       &                         \\
		Fr\'{e}chet ($\alpha$)    &          $1-\exp{\left(-z^{-\alpha}\right)},~~~z>1;\alpha>0$          &   $\frac{1}{\alpha}$    \\ \bottomrule
	\end{tabular}
	\label{Dist}
\end{table}
With regard to the proportion of censoring in the right tail, we consider three values: 0.10 (small), 0.35 (medium) and 0.65 (large). This allows us to study the performance of the estimators as censoring increases or decreases. 

\subsection{Simulation Design}\label{comp_EVIQp}

In this section, we examine the procedure for measuring the performance of point and interval estimators of the EVI. In the case of point estimators, the median of $R~(R=1000)$ repetitions was used as the point estimate of $\gamma_1,$ and MAD and median bias are obtained as the performance measures. 

On the other hand, the comparison of the confidence intervals are based on two properties: interval length and coverage probability. Before, we introduce the simulation algorithm to compute the diagnostics of the confidence interval, we present a procedure known as the conditional block bootstrap for obtaining samples for extreme value analysis in the case of censoring.

\subsubsection{Conditional Block Bootstrap for Censored Data}\label{cond.bootstap}
In order to obtain the performance measures, coverage probability and average interval length, the bootstrap samples are required. However, as stated in Section \ref{EVTCEN_genEst}, two scenarios in EVT in the case of censoring are to be avoided in this study. Firstly, if none of the observations are censored (i.e. as can happen in cases where $\gamma_1>0$  and $\gamma_2<0$), then the classical EVT estimation techniques apply. This has been widely studied in the literature and is not of interest in this paper. Secondly, for a completely censored case (which can occur when $\gamma_1<0$  and $\gamma_2>0$) the estimation of the EVI and the other extreme events are impossible. 
Therefore, any bootstrap procedure implemented for the estimation of parameters of extreme events for censored data must be constrained to exclude the above scenarios, particularly where the estimation is impossible. However, the bootstrap sampling \citet{Efron1993} and bootstrap for censored data \citet{Efron1981} do not guarantee the exclusion of these two scenarios. 

We present here a bootstrap procedure, termed the ``conditional block bootstrap", for selecting bootstrap samples that exclude the two scenarios in statistics of extremes when data is subject to random censoring. The conditional block bootstrap is a combination of ideas from the moving block bootstrap \citep{Efron1993} and the bootstrap for censored data \citep{Efron1981}. 

In this procedure, the censored data is grouped into randomly chosen blocks and it is crucial that each block must contain at least one censored observation. This ensures that the second case is eliminated from each generated bootstrap sample. The bootstrap observations are obtained by repeatedly sampling with replacement from these blocks and placing them together to form the bootstrap sample. Enough blocks must be sampled to obtain approximately the same sample size as the original censored sample. 

Given a sample of size, $n,$ a proportion of censoring in the right tail, $\wp,$ and assuming $\wp\le 0.5,$ the conditional block bootstrap procedure is as follows: 

\begin{enumerate}
	
	\item Group the $n$ observations into two groups namely, censored and noncensored with sample sizes $n_c$ and $n_{\bar{c}}$ respectively. Thus, $\wp=n_{\bar{c}}/n.$
	
	\item Let $d~(d\ge1)$ denote the number of censored observations to be included in each block. The size of each block, $s,$ is obtained as $(n\times d)/n_c.$ If $s$ is not an integer, then let $s=\lceil(n\times d)/n_c\rceil.$
	
	\item The number of blocks, $m,$ is chosen such that $n\approxeq m\times s.$ In the case, $n=m\times s,$ the blocks will have the same number of observations. Otherwise, if $n\approx m\times s,$ then $m$ is taken as $\lceil n/s\rceil,$ in which case the first $m-1$ blocks are allocated $s$ observations each and the remaining  $n-s(m-1)$ observations, allocated to the $m$th block. 
	
	\item  Let $b_i,~i=1,\ldots,m$ denote the $m$ blocks. Assign observations to each block by randomly sampling, $s-d$ observations without replacement from the noncensored group. In addition, randomly sample $d$ observations without replacement from the censored-group and assign to each block $b_i,~i=1,\ldots,m.$ Thus, each block would contain $d$ and $s-d$ observations that are censored and noncensored respectively.
	
	\item Sample $m$ times with replacement from $b_1, b_2,\ldots, b_m$ and place them together to form the bootstrap sample. Note that, more than $m$ blocks may be sampled, in the case, $n\approx m\times s,$ for the bootstrap sample to be approximately equal to the original sample size, $n.$ 
	
	\item Repeat (5) a large number of times, $B,$ to obtain $B$ bootstrap samples.
	
\end{enumerate}

In the case, $\wp< 0.5,$ the above procedure can be used to constitute the blocks. However, the allocations should be done such that each block contains at least one noncensored observation.

\subsubsection{Simulation algorithm}
The following algorithm is used to obtain performance measures of the estimators of $\gamma_1:$
\begin{enumerate}[\hspace{0.5cm}\textbf{A}1.]
	
	\item Generate $n$ observations from $Y$ and $C$ respectively, and hence, obtain $Z^{(1)}=\{Z_1,\ldots,Z_n\}$ and $\delta^{(1)}=\{\delta_1,\ldots,\delta_n\}.$ Repeat a large number of times $R-1~ (R=1000)$ to obtain $R$ pairs of $(Z^{(i)}, \delta^{(i)}),~i=1,\ldots,R$ samples.  
	
	\item Select the pair of samples,  $(Z^{(1)}, \delta^{(1)}).$ Draw $B~(B=1000)$ bootstrap samples each of size $n$ using the conditional block bootstrap procedure in Section \ref{cond.bootstap}. \label{CP: Step 2}
	
	\item  Compute the bootstrap replicates,  $\hat{\gamma}^{*(c,.)}_{1,1}, \ldots, \hat{\gamma}^{*(c,.)}_{1,B},$ using the estimators of $\gamma_1.$ \label{CP: Step 3}
	
	\item Compute the $100(1-\alpha)\%,$ bootstrap confidence interval. \label{CP: Step 4}
	
	\item Repeat  A\ref{CP: Step 2} through to A\ref{CP: Step 4} for the remainder of the pairs of samples,  $(Z^{(j)}, \delta^{(j)}), ~j=2, \ldots, R$ to obtain $R$ confidence intervals for $\gamma_1.$  \label{CP: Step 5}
	
	\item Compute the properties of confidence intervals i.e. coverage probability and average interval length using the $R$ confidence intervals in A\ref{CP: Step 5}.
	
\end{enumerate}

\subsection{Results and Discussions}

In this section, we discuss the results of the simulation study for each distribution. General comments across the various distribution are presented in the last section. The simulation results for the Burr, Pareto and Fr\'{e}chet distributions are presented in Appendices A, B and C respectively. In most cases, estimators having small values of MAD and median bias generally give better coverage probability and interval length. Therefore,  our performance measuring criterion focuses on the coverage probability (CP) and interval lengths. Generally, we regard a good estimator as having a coverage probability of at least 0.90 and a reasonable interval length among these estimators.

\subsubsection{Burr Distribution}
\begin{itemize}
	\item \textbf{For $\gamma_1=0.1:$}
	
	The ERM estimator is undoubtedly the best confidence interval estimator of $\gamma_1=0.1$ as it has small bias, MAD, CP approximately equal to the nominal level and shorter average confidence interval length. For percentage of censoring in the right tail, $\wp=10\%$ (or more generally $\wp\le 10\%$), other estimators of $\gamma_1=0.1$ including MOM, PMom and occasionally POT.L, have good CP values. However, these estimators have wider average interval lengths compared to the ERM estimator. Moreover, in the case of $\wp>10\%,$ ERM is the only estimator that has coverage probability close to the nominal level and has a shorter confidence interval length. Also, POT.L has good CP values but larger interval lengths, and hence, not recommended for estimating $\gamma_1=0.1.$ The apparent poor performance of most of the estimators of $\gamma_1$ may be due to the second-order parameter $\rho \to 0.$
	
	\item \textbf{For $\gamma_1=0.5:$}
	
	Hill, MVRB, Zipf, WW.KM and WW.L are the best estimators for intervals for small percentage of censoring less than or equal to 10\%. These estimators have CP values close to the nominal level and small average interval lengths. As the percentage of censoring increases, the MomR, ERM and POT.L estimators have the best CP values: the other estimators have poor coverage. In the case of large $\wp$ values, ERM and POT.L  are the top two estimators of $\gamma_1=0.5.$ Overall, ERM and POT.L are the estimators which have good CP values and can be considered for estimating $\gamma_1=0.5.$ However, POT.L has wider interval lengths and may not be appropriate for estimating $\gamma_1=0.5.$
	
	\item  \textbf{For $\gamma_1=0.9:$}
	
	Most of the confidence interval estimators perform very well for the estimation of $\gamma_1=0.9$ compared with $\gamma_1 \le 0.5.$ The Hill, MVRB, WW.L, ERM and POT.L estimators generally give CP values close to the desired level of 0.95 regardless of the percentage of censoring in the right tail. Among these estimators, POT.L has the largest average interval length followed by ERM. In addition, WW.KM and MomR are much better than the preceding estimators in terms of the average confidence interval lengths. In particular, the MomR estimator provides the best estimator of the EVI when there is heavy censoring: it has smallest interval length compared the estimators having CP values of approximately 0.95. However, its CP is less good at lower censoring. 
\end{itemize}

\subsubsection{Pareto Distribution}

\begin{itemize}
	
	\item \textbf{For $\gamma_1=0.1:$}
	In this case, regardless of the percentage of censoring in the right tail, few estimators of  $\gamma_1$ have CP values close to the nominal level and moderate interval lengths. These include UH, MOM, PMom and POT. The rest of the estimators have poor CP values close to zero except ERM and POT.L. However, POT.L has larger interval length, and hence, may not be appropriate an appropriate estimator of $\gamma_1.$ Thus, UH, MOM, PMom and POT are the most robust to censoring when estimating $\gamma_1=0.1.$

	\item  \textbf{For $\gamma_1=0.5:$}
	
	In the case of the estimation of $\gamma_1=0.5,$ more estimators satisfy the CP-Interval length criterion when compared to $\gamma_1=0.1.$ Estimators such as UH, ERM, MOM, PMom, POT and POT.L mostly have high CP values close to 0.95 regardless of the value of $\wp.$ Again, the POT.L estimator has the largest interval length. Overall, the MOM and ERM are the preferred estimators as they have better CP values and moderate interval lengths compared with the others. 
	
	\item  \textbf{For $\gamma_1=0.9:$}
	
	For small percentage of censoring in the right tail, $\wp=10\%,$ most of the estimators have good CP values. The exceptions to this include UH, PMom, WW.KM and WW.L. Also, when $\wp=0.35$ and $0.65$ the WW.KM, MOM, POT, POT.L and ERM estimators have good CP values and relatively moderate interval lengths.  However, POT.L always has the largest interval length of at least twice the estimator with the shortest interval length. Therefore, ERM, MOM and POT can be considered as more robust for the estimation of $\gamma_1=0.9,$ as $\wp$ increases.

\end{itemize}

\subsubsection{Fr\'{e}chet Distribution}

\begin{itemize}
	\item \textbf{For $\gamma_1=0.1:$}
	
	In the estimation of $\gamma_1=0.1,$ for small percentage of censoring, $\wp\le10\%,$ several confidence interval estimators with the exception of POT.L and UH provide good coverage probabilities and reasonable interval lengths. Among these estimators, Hill, MVRB, WW.L, WW.KM and ERM have CP values close to 0.95. In addition, for $\wp\ge0.35,$ similar performance is observed as with $\wp\le10.$ Here, we noticed a better performance in CP values of WW.L compared with WW.KM. This is in conformity to the simulation results reported in \citet{Worms2014}. Generally, the Hill, MVRB and ERM are the most appropriate for estimating $\gamma_1=0.1$ for various levels of censoring in the right tail.

	\item  \textbf{For $\gamma_1=0.5:$}
	
	At 10\% censoring in the right tail, the ERM, POT.L, POT, MomR and Zipf estimators provide good coverage probabilities. In terms of interval length, Zipf and MomR provide approximately half of the average interval lengths of the other estimators. Thus, these two estimators are the most appropriate estimators of $\gamma_1=0.5.$ However, as the percentage of censoring in the right tail increases, the ERM, POT.L and MOM estimators provide the best CP values. Moreover, the POT.L estimator has larger interval lengths, and hence, the ERM estimator is regarded as the most appropriate for estimating $\gamma_1=0.5.$

	\item   \textbf{For $\gamma_1=0.9:$}
	
	In the case of $\wp=10\%,$ most of the estimators of $\gamma_1$ performed well with CP values close to the the nominal level of 0.95  except Hill, MVRB, WW.KM and Zipf. The ERM, POT.L, POT, and PMom estimators consistently have CP values close to 0.95 and relatively good interval lengths. In addition, as with the case $\wp=10\%,$  the estimators of $\gamma_1=0.9$ exhibited similar performance when $\wp$ was increased to 35\% or 65\%. Overall, ERM, MOM and POT can be used as estimators of $\gamma_1=0.9$ that are more robust to censoring.

\end{itemize}

\subsubsection{General Comments}

As may be expected, no single estimator is universally the best for estimating the EVI across distributions, size of the EVI and percentage of censoring in the right tail. However, some common underlying behaviours exist. In what follows, we present some general comments on the estimators in all the distributions considered.

In the first place, we found that the estimators' performance diminish with increasing levels of the percentage of censoring. In this regard, we noticed either a decline in the values of the coverage probability or a wider confidence interval lengths as the percentage of censoring in the right tail increases.

Secondly, most estimators exhibit large bias when estimating small values of $\gamma_1,$ especially in the Burr and Pareto distributions. However, the proposed ERM estimator is an exception to this as it exhibits high coverage even for the Burr distribution.

Thirdly, in the case of specific distributions, the following observations were made. In the Burr distribution, ERM, MOM and MomR  are generally the best estimators of the EVI. For samples from the Fr\'{e}chet distribution, ERM and MOM are universally good for estimating various sizes of the EVI and most robust to censoring whereas in the case of samples from the Pareto distribution, ERM, PMom and POT  estimators of the EVI appear to be the best. 

Lastly, we found the two estimators, ERM and MOM as the most appropriate for the estimation of the EVI across all the distributions. In addition, these estimators are the most robust to censoring and the size of the EVI. More importantly, the proposed ERM estimator was observed to be consistently robust for the estimation of the EVI regardless of latter's size and the percentage of censoring. Moreover, the estimation of from the exponential regression, the basis of the ERM estimator, leads to estimators of the second order parameters. These second-order parameters can be used to obtain reduced-bias estimators of quantiles and exceedance probabilities.

\section{Practical Application}\label{sec_prac}

In this section, we present an application of the estimators of the EVI discussed in the previous section to study the tails of the distribution of the survival time of AIDS patients. Data was obtained from \citet{Venables2002} based on a study by Dr. P. J. Solomon and the Australian National Centre in HIV Epidemiology and Clinical Research. 

The data consists of 2,843 patients of which 1,761 patients died while the remaining were right censored. Out of the total number of patients, 2,754 were males, of which 1,708 died and the remaining 1,046 were right censored. In this study, we consider the male patients only. 

This data has been studied in the extreme value theory literature in \citet{Einmahl2008} and \citet{Ndao2014}. In the former, the EVI is used to assess the tail heaviness of the right tail of the survival function, $1-F,$ and extreme quantiles are estimated to obtain an indication of how long a healthy man can survive AIDS. The latter uses survival time as a response variable with the age of the patient at diagnosis as covariate to obtain conditional EVI (or tail index) and extreme quantiles. Thus, the tails of the distribution of the survival time of male AIDS patients is studied conditional on the age at diagnosis.

\begin{figure}[htp!]
	\centering
	
	\subfloat[]{%
		\includegraphics[height=7cm,width=.48\textwidth]{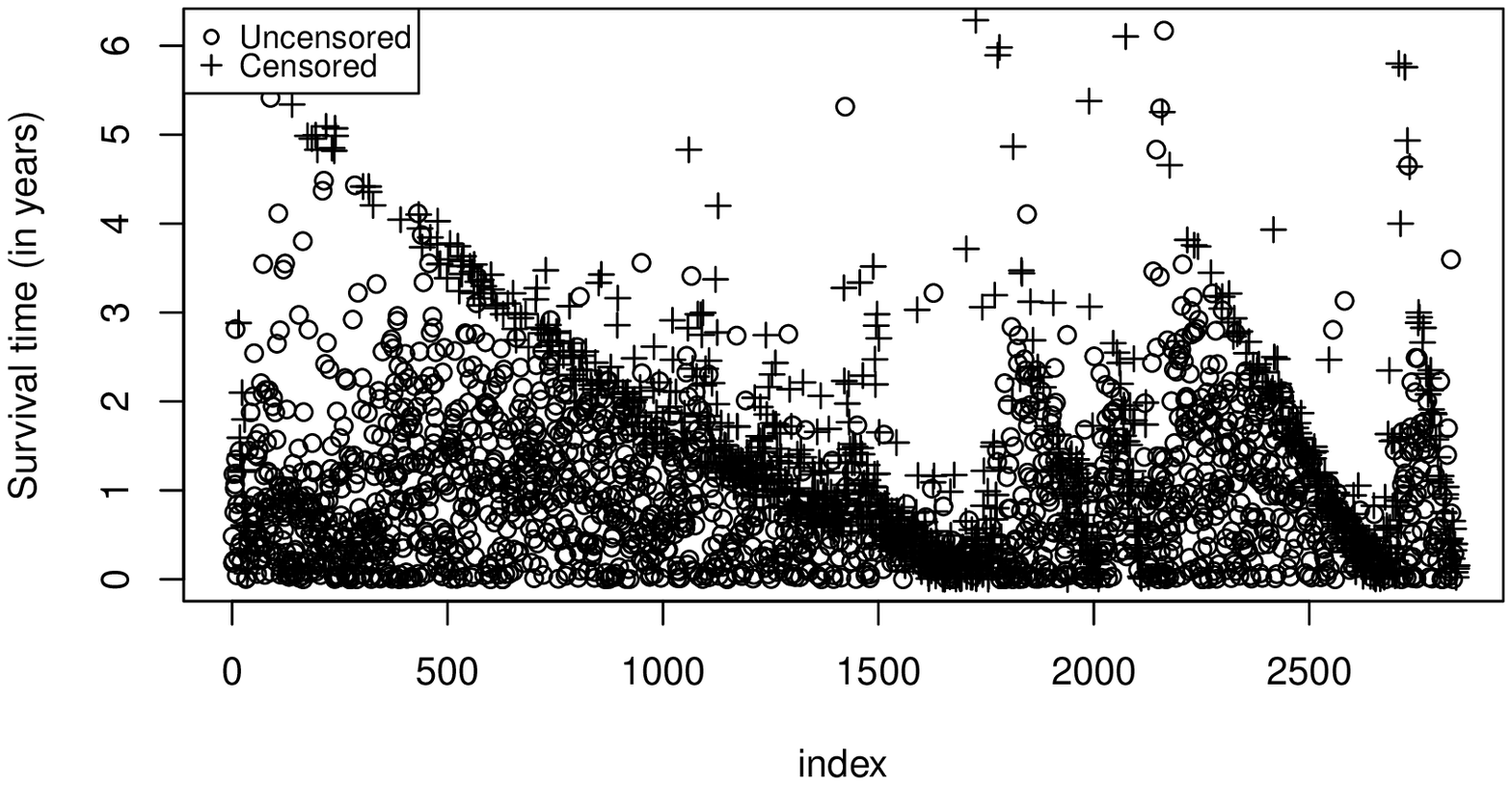}}\hfill
	\subfloat[]{%
		\includegraphics[height=6cm,width=.48\textwidth]{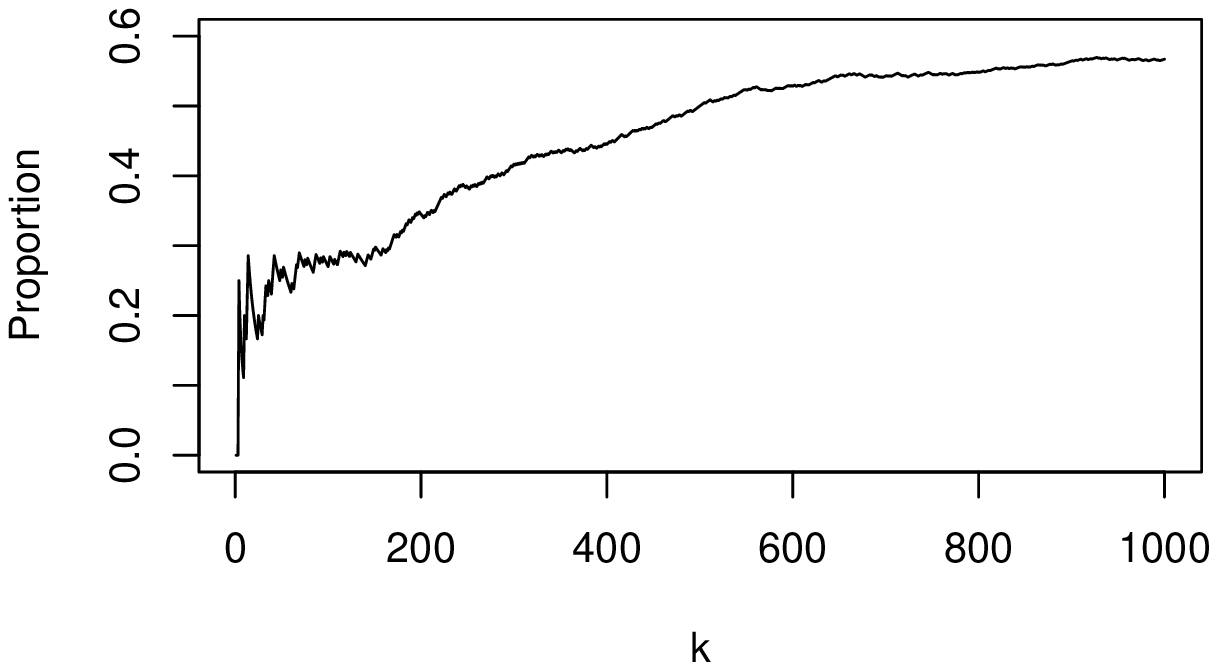}}\hfill
	\caption{ (a) Survival time of AIDS patients. (b) Estimates of the $\wp.$ }
	\label{AIDS}
\end{figure}

Figure \ref{AIDS} shows the scatter plot and histogram of the  Australian AIDS survival data. The scatter plot indicates that most of the males who survive longer are censored and the histogram indicates that there is a lower chance of survival after 7 years of diagnosis with AIDS.  

\begin{figure}
	
	\centering
	\subfloat[]{%
		\includegraphics[height=6cm,width=0.48\linewidth]{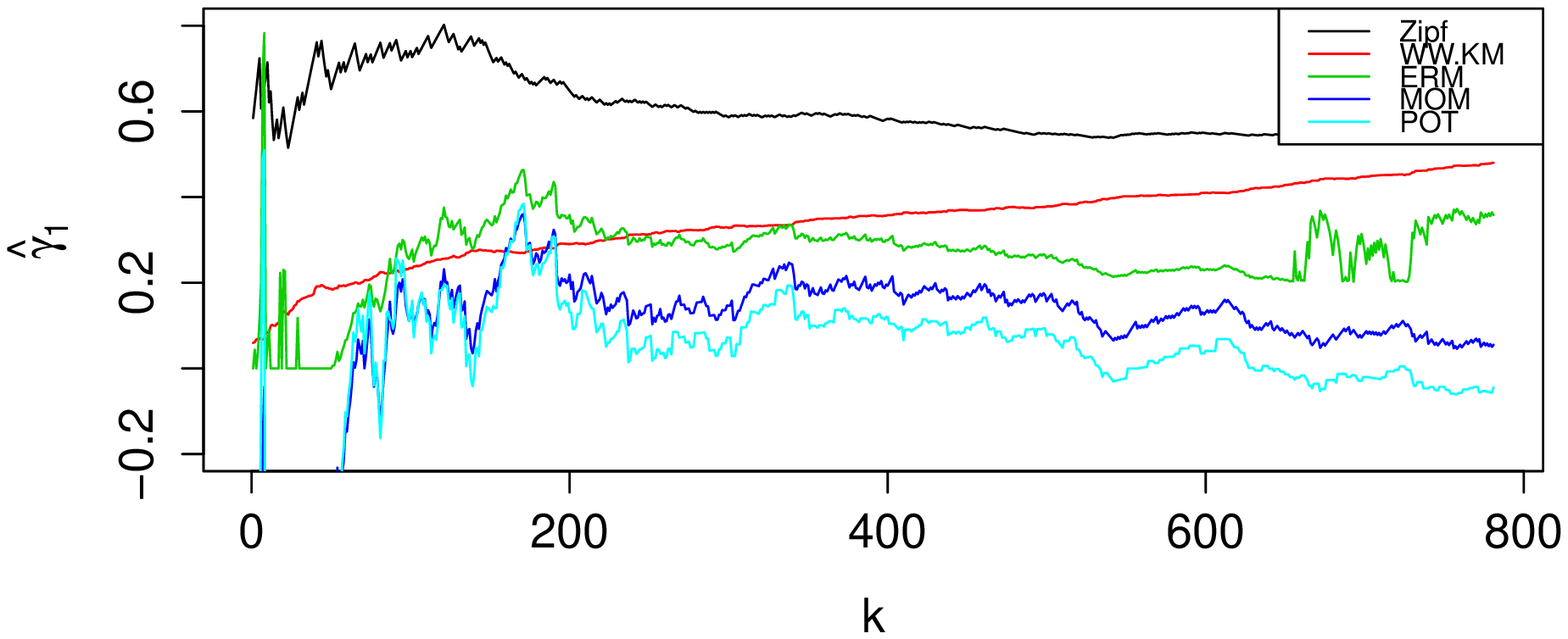}}\hfill
	\subfloat[]{%
		\includegraphics[height=6cm,width=.48\textwidth]{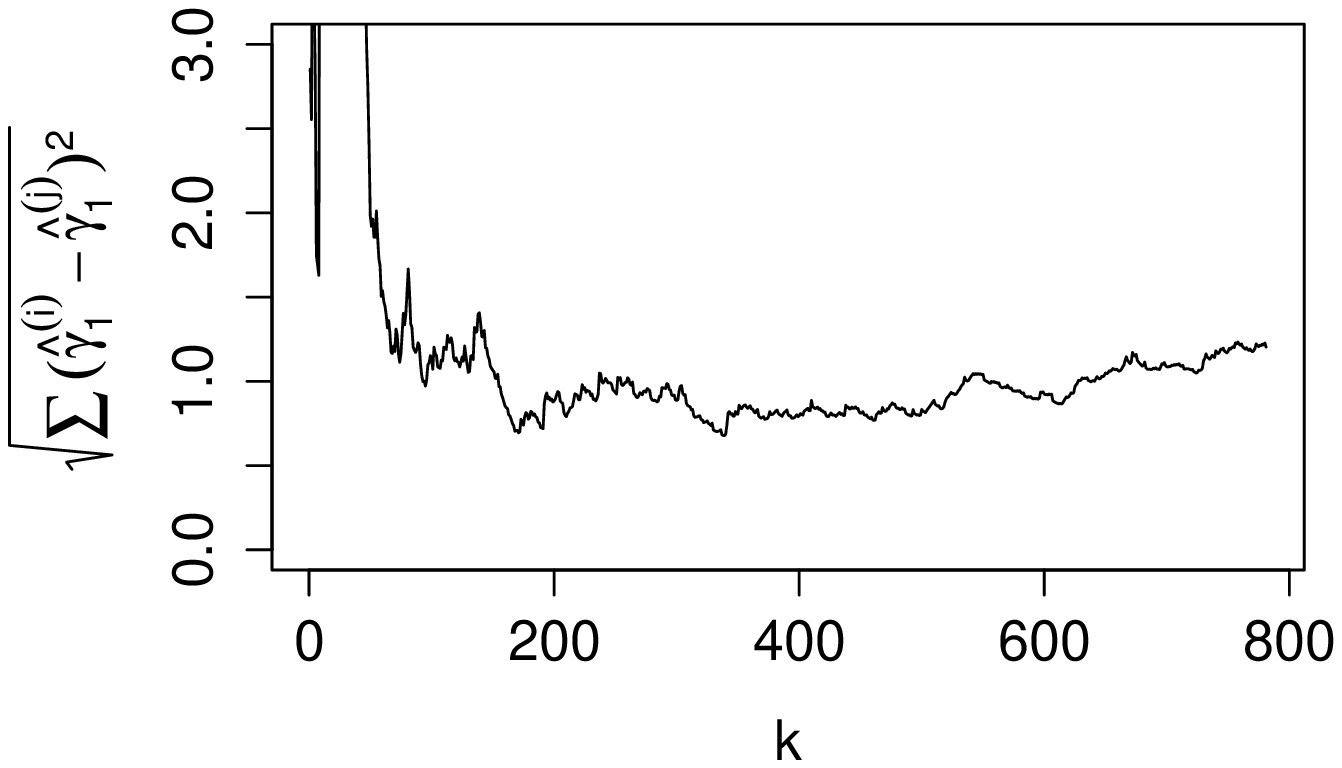}}\hfill
	\caption{Estimates of $\gamma_1,$ left panel; Heuristic choice of the threshold $k,$ right panel}
	\label{SurvEVIQp}

\end{figure}

The estimation of the EVI has been shown in the simulation to be sensitive to the value of $\wp.$ The values of $\wp$ must be reasonably moderate in the top order statistics to enable the application of the estimators of the EVI. Therefore, it is necessary in applications to assess the percentage (or proportion) of censoring in the right tail. The left panel of Figure \ref{SurvEVIQp} shows a plot of the proportion of censoring as a function of $k.$ \citet{Einmahl2008} chose the proportion of censoring as $\wp=0.28$ and justified the selection as corresponding to the most stable part of the graph i.e. $60\le k\le200.$ However, owing to the sensitivity of the estimators of $\gamma_1$ to $\wp,$ we compute our estimates using the actual $\wp$ values in the data. 

From the conclusions drawn from the simulation study and in order to make it less cumbersome, we selected five estimators for illustration. These estimators are ERM, POT, MOM, WW.KM and Zipf. The estimators of the EVI, $\gamma_1,$ are presented in the right panel of Figure \ref{SurvEVIQp}. As with the UH estimator used in \citet{Einmahl2008}, the estimators of $\gamma_1$  are relatively constant for $k\ge 200.$ 

Also, in practice, when a set of EVI estimators are to be taken into account,
\citet{Henriques2011} provide a simple heuristic approach to aid in selecting an appropriate threshold. 
We follow a modification of the heuristic approach of selecting an optimal $k$ instead of a percentage of the sample size as used in Section 3. Let $\gamma_1^{(i)},~i\in\Omega$ be the list of estimators under consideration where $\Omega=\{\mbox{Zipf, WW.KM, ERM, MOM, POT} \}.$ The optimum value of $k,$ is chosen as  

\begin{equation}\label{k_min}
k_{\mbox{opt}}=\underset{k}{\mathrm{argmin}}\sqrt{\sum_{(i,j)\in\Omega,~i\ne j}^{}\left(\hat{\gamma}_1^{(i)}-\hat{\gamma}_1^{(j)}\right)^2}.
\end{equation}

We apply (\ref{k_min}) to the EVI estimators for the AIDS survival data and the results are presented in the right panel of \ref{SurvEVIQp}. A closer look at the graph shows a stable region between 200 and 600: we choose $k_{\mbox{opt}}=339$ (which is equal to 12\% of the sample size and close to the 10\% used in the simulation study) for the estimation of $\gamma_1.$ 

The EVI estimates at $k_{\mbox{opt}}=339$ are shown in Table \ref{SurvTab}. In \citet{Einmahl2008}, only the generalised Hill estimator, $\hat{\gamma}_1^{(c,UH)}$  was used for the estimation of the EVI. The estimate of $\hat{\gamma}_1^{(c,UH)}$ was found to be 0.14. In addition, \citet{Ndao2014} estimates $\gamma_1$ as 0.304, 0.340 and 0.323 for males diagnosed with AIDS at ages 27, 37 and 47 years respectively. 

Therefore, with the exception of the Zipf estimator, all the other estimators considered give estimates within the range of the values provided by \citet{Einmahl2008} and \citet{Ndao2014}. In particular, our ERM estimator of $\gamma_1$ and the WW.KM give estimates close to that of \citet{Ndao2014}, although age was not considered as a factor. Moreover, the ERM estimator is quite stable for most part of the values of $k.$

\begin{table}[htp!]
	\centering
	\caption{Estimates of the EVI and the corresponding extreme quantile at $k_{opt}$}
	\begin{tabular}{llllll}
		\toprule
		&\multicolumn{5}{c}{EVI} \\\cmidrule{2-6}
		Estimator & WW.KM & Zipf  &  MOM &  POT&    ERM  \\
		Estimate  &0.334 &  0.587&  0.244&  0.193 & 0.334 \\\bottomrule
	\end{tabular}
	\label{SurvTab}
\end{table}

\section{Conclusions}\label{sec_conc}

This paper reviews various estimators of extreme value index when observations are subject to right random censoring. In addition, an estimator based on exponential regression model was proposed among others. Since the asymptotic distributions are not known for all the estimators, theoretical comparison was not possible. Therefore, a simulation study was conducted to compare the performance of the various estimators under different distribution, size of the EVI and percentage of censoring in the right tail. The performance criterion used were bias, MAD, confidence interval length and coverage probability.  The simulation results show that the performance of the estimators differ, depending on: the undelying distribution; EVI size; and percentage of censoring in the right tail. Therefore, no estimator was shown to be universally the best across all these scenarios. However, certain estimators perform reasonably well across most distributions. These are the estimators that we recommend as appropriate for the estimation of the EVI. In this regard, if a practitioner is interested in estimators that perform well across distributions in the sense of having good coverage and small interval size, then we recommend the proposed ERM and MOM estimators. The estimators that performed well in the simulation study were illustrated using real data on the survival of AIDS patients. Generally, we recommend that practitioners should assess the distribution of a dataset, size of $\gamma_1$ and proportion of censoring using other external information. This includes graphical plots to assist in knowing the tail behaviour of the underlying distribution and plot of the proportion of censoring at different values of $k.$ In addition, several estimators can be used to compute estimates of $\gamma_1$ to assess the possible size of $\gamma_1,$ and hence, the selection of an appropriate estimator. We believe that the findings from this simulation will help practitioners in the selection of estimators of EVI when data is subject to right random censoring.

\renewcommand{\bibname}{References}
\bibliographystyle{apalike}
\nocite{*}
\bibliography{Cens}
\appendix
	\addtocontents{toc}{\protect\setcounter{tocdepth}{0}}
	
	\section{Burr Distribution}
	
	\begin{table}[htp!]
		\centering
		\caption{Estimation of $\gamma_1=0.1$}
		\begin{adjustbox}{width=\linewidth}	
			\subfloat[$\wp=0.10$]{
				\begin{tabular}{lllll}
					\toprule
					Estimator&  MAD & MedBias & CP & $\bar{L}$ \\ 
					\hline
					Hill  & 0.13 & 0.13 & 0.00 & 0.13  \\ 
					MVRB  & 0.13 & 0.13 & 0.00 & 0.13  \\ 
					Zipf  & 0.12 & 0.12 & 0.01 & 0.12  \\ 
					UH  & 0.12 & -0.1 & 0.75 & 0.60  \\ 
					WW.KM  & 0.13 & 0.13 & 0.00 & 0.12  \\ 
					WW.L  & 0.13 & 0.13 & 0.00 & 0.13  \\ 
					MOM  & 0.14 & -0.13 & 0.86 & 0.75 \\ 
					MomR  & 0.10 & 0.10 & 0.01 & 0.10  \\ 
					PMom  & 0.18 & -0.17 & 0.82 & 0.82  \\ 
					POT  & 0.18 & -0.18 & 0.79 & 0.87 \\ 
					POT.L  & 0.95 & -0.95 & 0.52 & 3.17 \\ 
					ERM  & 0.06 & 0.04 & 0.97 & 0.28  \\ 
					\bottomrule
				\end{tabular}
			}\hfill
			\subfloat[$\wp=0.35$]{
				\begin{tabular}{llll}
					\toprule
					MAD & MedBias & CP & $\bar{L}$ \\ 
					\hline
					0.20 & 0.20 & 0.00 & 0.20  \\ 
					0.20 & 0.20 & 0.00 & 0.20  \\ 
					0.18 & 0.18 & 0.00 & 0.18  \\ 
					0.17 & -0.14 & 0.74 & 0.86  \\ 
					0.15 & 0.15 & 0.07 & 0.17\\ 
					0.18 & 0.18 & 0.00 & 0.16  \\ 
					0.21 & -0.19 & 0.85 & 1.09  \\ 
					0.16 & 0.16 & 0.00 & 0.16  \\ 
					0.25 & -0.25 & 0.82 & 1.19  \\ 
					0.27 & -0.26 & 0.76 & 1.20  \\ 
					0.40 & -0.38 & 0.72 & 3.11 \\
					0.08 & 0.06 & 0.98 & 0.36  \\ 
					\bottomrule
				\end{tabular}
			}\hfill
			\subfloat[$\wp=0.65$]{
				\begin{tabular}{llll}
					\toprule
					MAD & MedBias & CP & $\bar{L}$ \\ 
					\hline
					0.42 & 0.42 & 0.00 & 0.54  \\ 
					0.41 & 0.41 & 0.00 & 0.54 \\ 
					0.36 & 0.36 & 0.00 & 0.45  \\ 
					- 0.34 & -0.32 & 0.74 & 1.78  \\ 
					0.13 & 0.13 & 0.69 & 0.26  \\ 
					0.28 & 0.28 & 0.00 & 0.25  \\ 
					0.41 & -0.41 & 0.84 & 2.31 \\ 
					0.33 & 0.33 & 0.00 & 0.42 \\ 
					0.51 & -0.51 & 0.80 & 2.48 \\ 
					0.53 & -0.52 & 0.74 & 2.42  \\ 
					0.94 & -0.67 & 0.90 & 8.94  \\
					0.15 & 0.15 & 0.98 & 0.61  \\ 
					\bottomrule
				\end{tabular}
			}
		\end{adjustbox}		
	\end{table}

	\begin{table}[htp!]
		\centering
		\caption{Estimation of $\gamma_1=0.5$}
		\begin{adjustbox}{width=\linewidth}	
			
			\subfloat[$\wp=0.10$]{
				\begin{tabular}{lllll}
					\toprule
					Estimator & MAD & MedBias & CP & $\bar{L}$ \\ 
					\hline
					Hill  & 0.06 & 0.04 & 0.97 & 0.31 \\ 
					MVRB & 0.06 & 0.04 & 0.97 & 0.31 \\ 
					Zipf  & 0.07 & 0.06 &0.92& 0.33 \\ 
					UH  & 0.13 & -0.04 & 0.68 & 0.57 \\ 
					WW.KM  & 0.06 & 0.03 & 0.96 & 0.30 \\ 
					WW.L  & 0.06 & 0.04 & 0.94 & 0.31 \\ 
					MOM  & 0.12 & -0.06 & 0.83 & 0.66 \\ 
					MomR  & 0.07 & 0.00 & 0.91 & 0.29 \\ 
					PMom & 0.16 & -0.10 & 0.83 & 0.83 \\ 
					POT  & 0.16 & -0.08 & 0.91 & 1.01 \\ 
					POT.L  & 0.23 & -0.12 & 0.89 & 1.80 \\
					ERM  & 0.13 & -0.02 & 0.94 & 0.81 \\ 
					\bottomrule
				\end{tabular}
			}\hfill
			\subfloat[$\wp=0.35$]{
				\begin{tabular}{llll}
					\toprule
					MAD & MedBias & CP & $\bar{L}$ \\ 
					\hline
					0.08 & 0.06 & 0.91 & 0.39 \\ 
					0.07 & 0.06 & 0.92 & 0.39 \\ 
					0.09 & 0.08 & 0.90 & 0.41 \\ 
					0.16 & -0.06 & 0.75 & 0.84 \\ 
					0.07 & -0.01 & 0.89 & 0.34 \\ 
					0.07 & 0.05 & 0.91 & 0.36 \\ 
					0.17 & -0.10 & 0.84 & 0.96 \\ 
					0.06 & 0.02 & 0.93 & 0.36 \\ 
					0.23 & -0.15 & 0.84 & 1.16 \\ 
					0.24 & -0.14 & 0.91 & 1.35 \\ 
					0.31 & -0.14 & 0.92 & 3.00 \\
					0.14 & -0.02 & 0.95 & 0.90 \\ 
					\bottomrule
				\end{tabular}
			}\hfill
			\subfloat[$\wp=0.65$]{
				\begin{tabular}{llllll}
					\toprule
					& MedBias & CP & $\bar{L}$ \\ 
					\hline
					0.26 & 0.26 & 0.59 & 0.79 \\ 
					0.26 & 0.26 & 0.57 & 0.79 \\ 
					0.25 & 0.25 & 0.72 & 0.75 \\ 
					0.29 & -0.10 & 0.81 & 1.69 \\ 
					0.18 & -0.15 & 0.62 & 0.43 \\ 
					0.11 & 0.11 & 0.85 & 0.46 \\  
					0.30 & -0.19 & 0.86 & 2.08 \\ 
					0.18 & 0.18 & 0.81 & 0.68 \\ 
					0.37 & -0.27 & 0.84 & 2.34 \\ 
					0.38 & -0.26 & 0.88 & 2.45 \\ 
					0.67 & -0.20 & 0.96 & 6.76 \\
					0.18 & 0.06 & 0.94 & 1.17 \\ 
					\bottomrule
				\end{tabular}
			}
			
		\end{adjustbox}		
	\end{table}

	\begin{table}[htp!]
		\centering
		\caption{Estimation of $\gamma_1=0.9$}
		\begin{adjustbox}{width=\linewidth}	
			\subfloat[$\wp=0.10$]{
				\begin{tabular}{lllll}
					\toprule
					Estimator & MAD & MedBias & CP & $\bar{L}$ \\ 
					\hline
					Hill  & 0.10 & 0.01 & 0.95 & 0.54 \\ 
					MVRB  & 0.10 & 0.01 & 0.95 & 0.54 \\ 
					Zipf  & 0.13 & 0.07 & 0.88 & 0.59 \\ 
					UH  & 0.14 & -0.03 & 0.66 & 0.63 \\ 
					WW.KM  & 0.10 & -0.01 & 0.93 & 0.52 \\ 
					WW.L  & 0.10 & 0.01 & 0.95 & 0.54 \\ 
					MOM  & 0.14 & -0.05 & 0.82 & 0.74 \\ 
					MomR  & 0.12 & -0.03 & 0.85 & 0.52 \\ 
					PMom  & 0.21 & -0.09 & 0.81 & 1.00 \\ 
					POT  & 0.20 & -0.06 & 0.92 & 1.20 \\ 
					POT.L  & 0.20 & -0.06 & 0.93 & 1.60 \\
					ERM  & 0.24 & -0.03 & 0.94 & 1.24 \\ 
					\bottomrule
				\end{tabular}
			}\hfill
			\subfloat[$\wp=0.35$]
			{
				\begin{tabular}{llll}
					\toprule
					MAD & MedBias & CP & $\bar{L}$ \\ 
					\hline
					0.10 & 0.03 & 0.96 & 0.67 \\ 
					0.10 & 0.03 & 0.97 & 0.67 \\ 
					0.14 & 0.09 & 0.92 & 0.71 \\ 
					0.18 & -0.05 & 0.75 & 0.90 \\ 
					0.14 & -0.09 & 0.81 & 0.57 \\ 
					0.11 & 0.02 & 0.94 & 0.61 \\ 
					0.18 & -0.06 & 0.84 & 1.01 \\ 
					0.12 & -0.01 & 0.90 & 0.64 \\ 
					0.26 & -0.11 & 0.84 & 1.29 \\ 
					0.25 & -0.07 & 0.93 & 1.58 \\ 
					0.32 & -0.11 & 0.93 & 2.77 \\
					0.23 & 0.00 & 0.94 & 1.45 \\ 
					\bottomrule
				\end{tabular}
			}\hfill
			\subfloat[$\wp=0.65$]{
				\begin{tabular}{llll}
					\toprule
					MAD & MedBias & CP & $\bar{L}$ \\ 
					\hline
					0.19 & 0.17 & 0.93 & 1.08 \\ 
					0.19 & 0.16 & 0.93 & 1.10 \\ 
					0.22 & 0.20 & 0.92 & 1.14 \\ 
					0.33 & -0.07 & 0.81 & 1.81 \\ 
					0.41 & -0.41 & 0.35 & 0.59 \\ 
					0.15 & -0.02 & 0.91 & 0.70 \\ 
					0.34 & -0.12 & 0.86 & 2.06 \\ 
					0.18 & 0.09 & 0.95 & 1.02 \\ 
					0.42 & -0.22 & 0.85 & 2.43 \\ 
					0.42 & -0.20 & 0.91 & 2.79 \\ 
					0.64 & -0.21 & 0.96 & 6.80 \\
					0.29 & 0.01 & 0.96 & 1.89 \\ 
					\bottomrule
				\end{tabular}
			}
		\end{adjustbox}		
	\end{table}
	
	\newpage
	\section{Pareto Distribution}
	
	\begin{table}[htp!]
		\centering
		\caption{Estimation of $\gamma_1=0.1$}
		\begin{adjustbox}{width=\linewidth}	
			\subfloat[$\wp=0.10$]{
				\begin{tabular}{lllll}
					\toprule
					Estimator & MAD & MedBias & CP & $\bar{L}$ \\ 
					\hline
					Hill  & 0.31 & 0.31 & 0.00 & 0.16 \\ 
					MVRB  & 0.31 & 0.31 & 0.00 & 0.16 \\ 
					Zipf & 0.26 & 0.26 & 0.00 & 0.14 \\ 
					UH  & 0.08 & 0.06 & 0.94 & 0.38 \\ 
					WW.KM  & 0.30 & 0.30 & 0.00 & 0.15 \\ 
					WW.L  & 0.31 & 0.31 & 0.00 & 0.15 \\ 
					MOM  & 0.08 & 0.04 & 0.96 & 0.49 \\ 
					MomR & 0.24 & 0.24 & 0.00 & 0.13 \\ 
					PMom  & 0.09 & -0.03 & 0.96 & 0.57 \\ 
					POT  & 0.08 & -0.03 & 0.93 & 0.54 \\ 
					POT.L  & 0.76 & -0.76 & 0.67 & 3.31  \\ 
					ERM  & 0.12 & 0.12 & 0.87 & 0.34 \\ 
					\bottomrule
				\end{tabular}
			}\hfill
			\subfloat[$\wp=0.35$]{
				\begin{tabular}{llll}
					\toprule
					MAD & MedBias & CP & $\bar{L}$ \\ 
					\hline
					0.45 & 0.45 & 0.00 & 0.26 \\ 
					0.45 & 0.45 & 0.00 & 0.26 \\ 
					0.38 & 0.38 & 0.00 & 0.22 \\ 
					0.12 & 0.09 & 0.95 & 0.53 \\ 
					0.38 & 0.38 & 0.00 & 0.22 \\ 
					0.41 & 0.41 & 0.00 & 0.21 \\ 
					0.12 & 0.06 & 0.97 & 0.70 \\ 
					0.36 & 0.36 & 0.00 & 0.21 \\ 
					0.13 & -0.05 & 0.95 & 0.81 \\ 
					0.12 & -0.04 & 0.92 & 0.74 \\ 
					0.31 & -0.28 & 0.78 & 2.30 \\ 
					0.18 & 0.18 & 0.80 & 0.46 \\ 
					\bottomrule
				\end{tabular}
			}\hfill
			\subfloat[$\wp=0.65$]{
				\begin{tabular}{llll}
					\toprule
					MAD & MedBias & CP & $\bar{L}$ \\ 
					\hline
					0.88 & 0.88 & 0.00 & 0.65 \\ 
					0.88 & 0.88 & 0.00 & 0.66 \\ 
					0.73 & 0.73 & 0.00 & 0.55 \\ 
					0.20 & 0.15 & 0.95 & 1.05 \\ 
					0.40 & 0.40 & 0.01 & 0.46 \\ 
					0.63 & 0.63 & 0.00 & 0.34 \\ 
					0.23 & 0.07 & 0.97 & 1.43 \\ 
					0.70 & 0.70 & 0.00 & 0.52 \\ 
					0.23 & -0.10 & 0.95 & 1.65 \\ 
					0.23 & -0.09 & 0.92 & 1.40 \\ 
					1.43 & -1.35 & 0.83 & 4.76 \\ 
					0.34 & 0.34 & 0.83 & 0.79 \\ 
					\bottomrule
				\end{tabular}
			}
		\end{adjustbox}		
	\end{table}
	
	\begin{table}[htp!]
		\centering
		\caption{Estimation of $\gamma_1=0.5$}
		\begin{adjustbox}{width=\linewidth}	
			\subfloat[$\wp=0.10$]{
				\begin{tabular}{lllll}
					\toprule
					Esimator& MAD & MedBias & CP & $\bar{L}$ \\ 
					\hline
					Hill  & 0.16 & 0.16 & 0.26 & 0.26 \\ 
					MVRB  & 0.16 & 0.16 & 0.27 & 0.27 \\ 
					Zipf  & 0.13 & 0.13 & 0.57 & 0.27 \\ 
					UH  & 0.08 & 0.03 & 0.84 & 0.39 \\ 
					WW.KM  & 0.15 & 0.15 & 0.32 & 0.26 \\ 
					WW.L & 0.16 & 0.16 & 0.24 & 0.26 \\ 
					MOM  & 0.08 & 0.02 & 0.94 & 0.47 \\ 
					MomR  & 0.10 & 0.10 & 0.69 & 0.25 \\ 
					PMom  & 0.12 & -0.05 & 0.90 & 0.61 \\ 
					POT  & 0.11 & -0.03 & 0.93 & 0.66 \\ 
					POT.L  & 0.17 & -0.07 & 0.92 & 1.37 \\ 
					ERM & 0.10 & -0.00 & 0.94 & 0.66 \\ 
					\bottomrule
				\end{tabular}
			}\hfill
			\subfloat[$\wp=0.35$]{
				\begin{tabular}{llll}
					\toprule
					MAD & MedBias & CP & $\bar{L}$ \\ 
					\hline
					0.29 & 0.29 & 0.03 & 0.38 \\ 
					0.29 & 0.29 & 0.03 & 0.38 \\ 
					0.24 & 0.24 & 0.21 & 0.36 \\ 
					0.11 & 0.06 & 0.88 & 0.57 \\ 
					0.20 & 0.20 & 0.55 & 0.33 \\ 
					0.25 & 0.25 & 0.05 & 0.33 \\ 
					0.12 & 0.03 & 0.95 & 0.68 \\ 
					0.20 & 0.20 & 0.26 & 0.33 \\ 
					0.16 & -0.07 & 0.90 & 0.849 \\ 
					0.15 & -0.05 & 0.92 & 0.89 \\ 
					0.20 & -0.05 & 0.92 & 1.76 \\ 
					0.11 & 0.02 & 0.94 & 0.79 \\ 
					\bottomrule
				\end{tabular}
			}\hfill
			\subfloat[$\wp=0.65$]{
				\begin{tabular}{llll}
					\toprule
					MAD & MedBias & CP & $\bar{L}$ \\ 
					\hline
					0.70 & 0.70 & 0.00 & 0.82 \\ 
					0.70 & 0.70 & 0.00 & 0.83 \\ 
					0.57 & 0.57 & 0.01 & 0.75 \\ 
					0.23 & 0.13 & 0.93 & 1.09 \\ 
					0.16 & 0.11 & 0.91 & 0.57 \\ 
					0.43 & 0.43 & 0.00 & 0.49 \\ 
					0.25 & 0.08 & 0.94 & 1.45 \\ 
					0.52 & 0.52 & 0.01 & 0.70 \\ 
					0.27 & -0.10 & 0.90 & 1.69 \\ 
					0.26 & -0.09 & 0.91 & 1.55 \\ 
					0.41 & -0.09 & 0.96 & 3.94 \\ 
					0.25 & 0.21 & 0.95 & 1.13 \\ 
					\bottomrule
				\end{tabular}
			}
		\end{adjustbox}		
	\end{table}

	\begin{table}[htp!]
		\centering
		\caption{Estimation of $\gamma_1=0.9$}
		\begin{adjustbox}{width=\linewidth}	
			\subfloat[$\wp=0.10$]{
				\begin{tabular}{lllll}
					\toprule
					Estimator & MAD & MedBias & CP & $\bar{L}$ \\ 
					\hline
					Hill  & 0.09 & 0.08 & 0.90 & 0.40 \\ 
					MVRB  & 0.09 & 0.08 & 0.90 & 0.40 \\ 
					Zipf  & 0.10 & 0.08 & 0.91 & 0.44 \\ 
					UH  & 0.10 & 0.01 & 0.75 & 0.46 \\ 
					WW.KM  & 0.08 & 0.07 & 0.90 & 0.39 \\ 
					WW.L  & 0.09 & 0.08 & 0.88 & 0.40 \\ 
					MOM  & 0.10 & 0.00 & 0.89 & 0.53 \\ 
					MomR  & 0.08 & 0.02 & 0.92 & 0.41 \\ 
					PMom  & 0.15 & -0.06 & 0.85 & 0.73 \\ 
					POT  & 0.14 & -0.03 & 0.93 & 0.82\\ 
					POT.L  & 0.15 & -0.03 & 0.93 & 1.04 \\
					ERM  & 0.17 & -0.04 & 0.92 & 0.97 \\ 
					\bottomrule
				\end{tabular}
			}\hfill
			\subfloat[$\wp=0.35$]{
				\begin{tabular}{llll}
					\toprule
					MAD & MedBias & CP & $\bar{L}$ \\ 
					\hline
					0.19 & 0.19 & 0.70 & 0.52 \\ 
					0.19 & 0.19 & 0.71 & 0.52 \\ 
					0.17 & 0.17 & 0.83 & 0.55 \\ 
					0.14 & 0.05 & 0.85 & 0.64 \\ 
					0.11 & 0.08 & 0.93 & 0.51 \\ 
					0.16 & 0.16 & 0.77 & 0.49 \\ 
					0.14 & 0.04 & 0.92 & 0.73 \\ 
					0.12 & 0.10 & 0.88 & 0.51 \\ 
					0.19 & -0.04 & 0.88 & 0.95 \\ 
					0.19 & -0.02 & 0.93 & 1.05 \\ 
					0.21 & 0.00 & 0.94 & 1.67 \\
					0.19 & 0.01 & 0.93 & 1.13 \\ 
					\bottomrule
				\end{tabular}
			}\hfill
			\subfloat[$\wp=0.65$]{
				\begin{tabular}{llll}
					\toprule
					MAD & MedBias & CP & $\bar{L}$ \\ 
					\hline
					0.52 & 0.52 & 0.19 & 0.98 \\ 
					0.52 & 0.52 & 0.15 & 0.98 \\ 
					0.45 & 0.45 & 0.49 & 0.92 \\ 
					0.23 & 0.11 & 0.92 & 1.17 \\ 
					0.23 & -0.13 & 0.76 & 0.71 \\ 
					0.25 & 0.25 & 0.59 & 0.64 \\ 
					0.23 & 0.07 & 0.94 & 1.45 \\ 
					0.37 & 0.37 & 0.59 & 0.86 \\ 
					0.28 & -0.11 & 0.89 & 1.73 \\ 
					0.27 & -0.08 & 0.94 & 1.72 \\ 
					0.47 & -0.07 & 0.95 & 4.21 \\ 
					0.24 & 0.10 & 0.95 & 1.49 \\ 
					\bottomrule
				\end{tabular}
			}
		\end{adjustbox}		
	\end{table}
	
	
	\newpage
	\section{Fr\'{e}chet Distribution}
	
	\begin{table}[htp!]
		\centering
		\caption{Estimation of $\gamma_1=0.1$}
		\begin{adjustbox}{width=\linewidth}	
			\subfloat[$\wp=0.10$]{
				\begin{tabular}{lllll}
					\toprule
					Estimator &  MAD & MedBias & CP & $\bar{L}$ \\ 
					\hline
					Hill  & 0.01 & 0.01 & 0.96 & 0.06 \\ 
					MVRB  & 0.01 & -0.00 & 0.94 & 0.06 \\ 
					Zipf  & 0.02 & 0.01 & 0.90 & 0.07 \\ 
					UH  & 0.10 & -0.06 & 0.82 & 0.60 \\ 
					WW.KM & 0.01 & 0.00 & 0.96 & 0.06 \\ 
					WW.L  & 0.01 & 0.01 & 0.95 & 0.06 \\ 
					MOM  & 0.11 & -0.08 & 0.89 & 0.62 \\ 
					MomR & 0.01 & 0.00 & 0.87 & 0.06 \\ 
					PMom  & 0.12& -0.08 & 0.89 & 0.66 \\ 
					POT  & 0.11 & -0.05 & 0.97 & 0.84 \\ 
					POT.L & 0.44 & -0.97 & 0.57 & 3.32 \\
					ERM & 0.03 & 0.00 & 0.95 & 0.16 \\ 
					\bottomrule
				\end{tabular}
			}\hfill	\subfloat[$\wp=0.35$]{
				\begin{tabular}{llll}
					\toprule
					MAD & MedBias & CP & $\bar{L}$ \\ 
					\hline
					0.01 & 0.01 & 0.95 & 0.08 \\ 
					0.01 & 0.01 & 0.95 & 0.08 \\ 
					0.017 & 0.02 & 0.93 & 0.09 \\ 
					0.16 & -0.10 & 0.83 & 0.87 \\ 
					0.01 & -0.00 & 0.88 & 0.07 \\ 
					0.01 & 0.01 & 0.94 & 0.07 \\ 
					0.14 & -0.12 & 0.91 & 0.89 \\ 
					0.02 & 0.00 & 0.92 & 0.07 \\ 
					0.17 & -0.13 & 0.90 & 0.94 \\ 
					0.13 & -0.05 & 0.98 & 1.28 \\ 
					0.39 & -0.34 & 0.79 & 3.11 \\
					0.03 & -0.00 & 0.97 & 0.17 \\ 
					\bottomrule
				\end{tabular}
			}\hfill
			\subfloat[$\wp=0.65$]{
				\begin{tabular}{llll}
					\toprule
					MAD & MedBias & CP & $\bar{L}$ \\ 
					\hline
					0.02 & 0.02 & 0.91 & 0.13 \\ 
					0.02 & 0.02 & 0.90 & 0.13 \\ 
					0.03 & 0.03 & 0.91 & 0.14 \\ 
					0.31 & -0.16 & 0.85 & 2.03 \\ 
					0.05 & -0.05 & 0.37 & 0.07 \\ 
					0.07 & 0.00 & 0.94 & 0.08 \\ 
					0.31 & -0.20 & 0.86 & 2.12 \\ 
					0.02 & 0.01 & 0.94 & 0.13 \\ 
					0.32 & -0.21 & 0.86 & 2.16 \\ 
					0.23 & 0.07 & 0.97 & 4.04 \\ 
					1.06 & -0.67 & 0.94 & 8.48 \\
					0.03 & 0.01 & 0.97 & 0.21 \\ 
					\bottomrule
				\end{tabular}
			}
			
		\end{adjustbox}		
	\end{table}
	
	\begin{table}[htp!]
		\centering
		\caption{Estimation of $\gamma_1=0.5$}
		\begin{adjustbox}{width=\linewidth}	
			\subfloat[$\wp=0.10$]{
				\begin{tabular}{lllll}
					\toprule
					Estimator	&  MAD & MedBias & CP & $\bar{L}$ \\ 
					\hline
					Hill  & 0.03 & 0.02 & 0.83 & 0.10 \\ 
					MVRB  & 0.02 & 0.02 & 0.86 & 0.10 \\ 
					Zipf  & 0.02 & 0.02 & 0.92 & 0.12 \\ 
					UH  & 0.04 & -0.01 & 0.82 & 0.19 \\ 
					WW.KM  & 0.02 & 0.02 & 0.84 & 0.10 \\ 
					WW.L  & 0.02 & 0.02 & 0.83 & 0.10 \\ 
					MOM  & 0.03 & -0.01 & 0.92 & 0.20 \\ 
					MomR  & 0.02 & 0.01 & 0.94 & 0.12 \\ 
					PMom & 0.05 & -0.02 & 0.91 & 0.27 \\ 
					POT  & 0.05 & -0.02 & 0.93 & 0.28 \\ 
					POT.L  & 0.05 & -0.02 & 0.94 & 0.30 \\
					ERM  & 0.04 & -0.01 & 0.94 & 0.24 \\ 
					\bottomrule
				\end{tabular}
			}\hfill
			
			\subfloat[$\wp=0.35$]{
				\begin{tabular}{llll}
					\toprule
					MAD & MedBias & CP & $\bar{L}$ \\ 
					\hline
					0.04 & 0.04 & 0.70 & 0.12 \\ 
					0.04 & 0.04 & 0.70 & 0.12 \\ 
					0.04 & 0.03 & 0.87 & 0.13 \\ 
					0.05 & -0.03 & 0.79 & 0.25 \\ 
					0.03 & 0.02 & 0.93 & 0.13 \\ 
					0.04 & 0.04 & 0.79 & 0.12 \\ 
					0.05 & -0.03 & 0.90 & 0.27 \\ 
					0.03 & 0.02 & 0.91 & 0.13 \\ 
					0.08 & -0.05 & 0.88 & 0.34 \\ 
					0.07 & -0.05 & 0.70 & 0.36 \\ 
					0.08 & -0.04 & 0.93 & 0.83 \\ 
					0.04 & -0.01 & 0.95 & 0.25 \\ 
					\bottomrule
				\end{tabular}
			}\hfill
			\subfloat[$\wp=0.65$]{
				\begin{tabular}{llll}
					\toprule
					MAD & MedBias & CP & $\bar{L}$ \\ 
					\hline
					0.10 & 0.10 & 0.31 & 0.18 \\ 
					0.10 & 0.10 & 0.31 & 0.18 \\ 
					0.09 & 0.09 & 0.56 & 0.20 \\ 
					0.10 & -0.04 & 0.82 & 0.53 \\ 
					0.11 & -0.10 & 0.57 & 0.27 \\ 
					0.05 & 0.04 & 0.90 & 0.21 \\ 
					0.10 & -0.04 & 0.90 & 0.56 \\ 
					0.08 & 0.08 & 0.61 & 0.19 \\ 
					0.12 & -0.07 & 0.88 & 0.64 \\ 
					0.12 & -0.07 & 0.89 & 0.66 \\ 
					0.21 & -0.10 & 0.92 & 1.59 \\
					0.06 & 0.05 & 0.95 & 0.31 \\ 
					\bottomrule
				\end{tabular}
			}
		\end{adjustbox}		
	\end{table}
	
	
	\begin{table}[htp!]
		\centering
		\caption{Estimation of $\gamma_1=0.9$}
		\begin{adjustbox}{width=\linewidth}	
			\subfloat[$\wp=0.10$]{
				\begin{tabular}{lllll}
					\toprule
					Estimator	& MAD & MedBias & CP & $\bar{L}$ \\ 
					\hline
					Hill  & 0.05 & 0.05 & 0.85 & 0.17 \\ 
					MVRB  & 0.05 & 0.04 & 0.87 & 0.17 \\ 
					Zipf  & 0.05 & 0.04 & 0.92 & 0.22 \\ 
					UH  & 0.04 & 0.01 & 0.91& 0.23 \\ 
					WW.KM  & 0.04 & 0.04 & 0.86 & 0.17 \\ 
					WW.L  & 0.05 & 0.05 & 0.85 & 0.17 \\ 
					MOM  & 0.04 & 0.01 & 0.95 & 0.25 \\ 
					MomR  & 0.04 & 0.02 & 0.94 & 0.21 \\ 
					PMom  & 0.06 & -0.01 & 0.92 & 0.36 \\ 
					POT  & 0.06 & -0.01 & 0.97 & 0.36 \\ 
					POT.L  & 0.06 & -0.01 & 0.96 & 0.37 \\
					ERM  & 0.07 & -0.01 & 0.96 & 0.44 \\ 
					\bottomrule
				\end{tabular}
			}\hfill			\subfloat[$\wp=0.35$]{
				\begin{tabular}{llll}
					\toprule
					MAD & MedBias & CP & $\bar{L}$ \\ 
					\hline
					0.08 & 0.08 & 0.66 & 0.21 \\ 
					0.07 & 0.07 & 0.77 & 0.21 \\ 
					0.06 & 0.06 & 0.87 & 0.24 \\ 
					0.06 & 0.01 & 0.89 & 0.30 \\ 
					0.05 & 0.04 & 0.94 & 0.23 \\ 
					0.07 & 0.07 & 0.79 & 0.22 \\ 
					0.06 & 0.01 & 0.93 & 0.32 \\ 
					0.05 & 0.04 & 0.92 & 0.24 \\ 
					0.08 & -0.04 & 0.90 & 0.43 \\ 
					0.08 & -0.03 & 0.93 & 0.44 \\ 
					0.10 & -0.02 & 0.94 & 0.54 \\ 
					0.08 & -0.03 & 0.93 & 0.46 \\ 
					\bottomrule
				\end{tabular}
			}\hfill
			\subfloat[$\wp=0.65$]	{
				\begin{tabular}{llll}
					\toprule
					MAD & MedBias & CP & $\bar{L}$ \\ 
					\hline
					0.06 & 0.19 & 0.28 & 0.34 \\ 
					0.06 & 0.19 & 0.23 & 0.34 \\ 
					0.06 & 0.17 & 0.53 & 0.36 \\ 
					0.09 & 0.04 & 0.93 & 0.56 \\ 
					0.14 & -0.19 & 0.60 & 0.48 \\ 
					0.07 & 0.06 & 0.92 & 0.34 \\ 
					0.10 & 0.03 & 0.94 & 0.60 \\ 
					0.06 & 0.14 & 0.60 & 0.35 \\ 
					0.12 & -0.01 & 0.93 & 0.73 \\ 
					0.12 & -0.01 & 0.95 & 0.75 \\ 
					0.20 & -0.08 & 0.96 & 1.31 \\ 
					0.09 & 0.07 & 0.93 & 0.56 \\ 
					\bottomrule
			\end{tabular}	}
		\end{adjustbox}		
	\end{table}
\end{document}